\RequirePackage{tikz}
\usetikzlibrary{fit}
\usetikzlibrary{calc}
\documentclass{article}%

\usepackage{mathptmx}
\usepackage{mathrsfs}

\usepackage{tabularx} 
\usepackage{amssymb,amsfonts,amsmath}
\usepackage{amsthm}
\usepackage{nicefrac}
\usepackage{multirow}
\usepackage{subfig}
\usepackage{rotating}
\usepackage{cancel}
\usepackage[utf8]{inputenc}
\usepackage{threeparttable}
\usepackage{booktabs}
\usepackage{todonotes}

\usepackage{authblk}
\usepackage[a4paper, total={6in, 9in}]{geometry}

\newtheorem{theorem}{Theorem}%
\newtheorem{proposition}[theorem]{Proposition}%

\newtheorem{example}[theorem]{Example}%
\newtheorem{definition}[theorem]{Definition}%
\newtheorem{corollary}[theorem]{Corollary}
\newtheorem{observation}[theorem]{Observation}

\newenvironment{proofs}{\noindent{\bf Proof.}\hspace*{1em}}{\literalqed\medskip}

\newcommand{\proofonlyif}{\smallskip\textit{Only if:\quad}}
\newcommand{\proofif}{\smallskip\textit{If:\quad}}

\newcommand\qedblob{\mbox{\qed}}
\def\literalqed{{\ \nolinebreak\hfill\mbox{\qedblob\quad}}}

\newcommand{\succeqsf}{\succeq^{\mathit{sumSF}}}
\newcommand{\succsf}{\succ^{\mathit{sumSF}}}

\newcommand{\simsf}{\sim^{\mathit{sumSF}}}

\newcommand{\utilitysfsum}{u^{\mathit{sumSF}}}

\newcommand{\succeqsfmin}{\succeq^{\mathit{minSF}}}
\newcommand{\succsfmin}{\succ^{\mathit{minSF}}}
\newcommand{\simsfmin}{\sim^{\mathit{minSF}}}
\newcommand{\utilitysfmin}{u^{\mathit{minSF}}}

\newcommand{\succeqequal}{\succeq^{\mathit{sumEQ}}}
\newcommand{\succequal}{\succ^{\mathit{sumEQ}}}

\newcommand{\simequal}{\sim^{\mathit{sumEQ}}}

\newcommand{\utilityeqsum}{u^{\mathit{sumEQ}}}

\newcommand{\succeqequalmin}{\succeq^{\mathit{minEQ}}}
\newcommand{\succequalmin}{\succ^{\mathit{minEQ}}}
\newcommand{\simequalmin}{\sim^{\mathit{minEQ}}}
\newcommand{\utilityeqmin}{u^{\mathit{minEQ}}}

\newcommand{\succeqal}{\succeq^{\mathit{sumAL}}}
\newcommand{\succal}{\succ^{\mathit{sumAL}}}

\newcommand{\simal}{\sim^{\mathit{sumAL}}}

\newcommand{\utilityalsum}{u^{\mathit{sumAL}}}

\newcommand{\succeqalmin}{\succeq^{\mathit{minAL}}}
\newcommand{\succalmin}{\succ^{\mathit{minAL}}}
\newcommand{\simalmin}{\sim^{\mathit{minAL}}}
\newcommand{\utilityalmin}{u^{\mathit{minAL}}}

\newcommand{\friendsum}[2]{\mathrm{sum}^F_{#1}(#2)}
\newcommand{\friendmin}[2]{\mathrm{min}^F_{#1}(#2)}
\newcommand{\friendisum}[2]{\mathrm{sum}^{F+}_{#1}(#2)}
\newcommand{\friendimin}[2]{\mathrm{min}^{F+}_{#1}(#2)}

\newcommand{\friendsumprime}[2]{\mathrm{sum}^{F\prime}_{#1}(#2)}

\newcommand{\friendminextprime}[2]{\mathrm{min}^{F\prime}_{#1}(#2)}
\newcommand{\friendiminextprime}[2]{\mathrm{min}^{F+\prime}_{#1}(#2)}

\newcommand{\coalstr}{\mathcal{C}_{N}}

\newcommand{\xthreeclong}{\rm \mbox{\sc{}Exact Cover by 3-Sets}}

\newcommand{\rxthreec}{\ensuremath{\textsc{RX3C}}}

\newcommand{\pol}{\ensuremath{\mathrm{P}}}
\newcommand{\np}{\ensuremath{\mathrm{NP}}}

\newcommand{\co}{\ensuremath{\mathrm{co}}}
\newcommand{\conp}{\co\np}

\newcommand{\condition}{\,{\mid}\:}

\newcommand{\OMIT}[1]{} %
\newcommand{\OOMIT}[1]{} %

\newcommand{\EP}[3]{
	\smallskip
	\begin{center}
		{\small 
			\begin{tabularx}{1.0\columnwidth}{ll}
				\toprule
				\multicolumn{2}{c}{\sc{#1}} \\
				\midrule
				{\bf Given:}& \parbox[t]{0.83\columnwidth}{#2\vspace*{1mm}} \\%
				{\bf Question:}& \parbox[t]{0.83\columnwidth}{#3\vspace*{.5mm}} \\ 
				\bottomrule
			\end{tabularx}
		}
	\end{center}
	\smallskip
}

\begin{document}

\title{Altruism in Coalition Formation Games}

\author{Anna Maria Kerkmann}
\author{Simon Cramer}
\author{J{\"o}rg Rothe}

\affil{Heinrich-Heine-Universit{\"a}t D{\"u}sseldorf, Germany}

\maketitle

\begin{abstract}
	Nguyen \emph{et al.}~\cite{ngu-rey-rey-rot-sch:c:altruistic-hedonic-games}
	introduced altruistic hedonic games in which agents' utilities depend
	not only on their own preferences but also on those of their friends
	in the same coalition.  We propose to extend their model to
	coalition formation games in general, considering also the
	friends in other coalitions.  
	Comparing our model to altruistic hedonic games, we argue that
	excluding some friends from the altruistic behavior
	of an agent is a major disadvantage that 
	comes with the restriction to hedonic games.
	After introducing our model
	and showing some desirable properties,
	we 
	additionally
	study %
	some
	common stability notions %
	and
	provide a computational analysis of the associated verification and
	existence problems.
\end{abstract}

%\keywords{Coalition formation, Hedonic game, Altruism, Cooperative game theory}

\section{Introduction}
We consider coalition formation games where agents have to form coalitions based on their preferences.
Among other compact representations of hedonic coalition formation games,
Dimitrov \emph{et al.}~\cite{dim-bor-hen-sun:j:core-stability-hedonic-games}
in particular proposed
the \emph{friends-and-enemies encoding with friend-oriented preferences}
which involves 
a \emph{network of friends}: a (simple) undirected graph whose 
vertices are the players and where two players are connected by an edge exactly if they are friends of each other. Players not connected by an edge consider each other as enemies.
Under friend-oriented preferences,
player $i$ 
prefers a coalition $C$ to a coalition $D$ if
$C$ contains more of $i$'s friends than~$D$,
or $C$ and $D$ have the same number of $i$'s friends but 
$C$ contains fewer enemies of $i$'s than~$D$. 
This is a special case
of the \emph{additive encoding}
\cite{azi-bra-see:j:computing-desirable-properties-in-hedonic-games}.
For more background on these two compact representations,
see Section~\ref{sec:model} and the book chapter by
Aziz and Savani~\cite{azi-sav:b:handbook-comsoc-hedonic-games}.

Based on friend-oriented preferences,
Nguyen \emph{et al.}~\cite{ngu-rey-rey-rot-sch:c:altruistic-hedonic-games}
introduced \emph{altruistic hedonic games} 
where agents gain utility not only from their own satisfaction
but also from their friends' satisfaction.
However,
Nguyen \emph{et al.}~\cite{ngu-rey-rey-rot-sch:c:altruistic-hedonic-games}
specifically considered hedonic games only,
which require that an agent's utility only depends on her own
coalition.  In their interpretation of altruism, the utility of an
agent is composed of the agent's own valuation of her coalition
and the valuation of all this agent's friends \emph{in this coalition}.
While
Nguyen \emph{et al.}~\cite{ngu-rey-rey-rot-sch:c:altruistic-hedonic-games}
used the average when aggregating 
some agents' valuations,
Wiechers and Rothe~\cite{wie-rot:c:stability-in-minimization-based-altruistic-hedonic-games}
proposed a variant of altruistic hedonic games 
where
some agents' valuations are aggregated by taking the minimum.

Inspired by the idea of altruism, we extend
the model of altruism in hedonic games to coalition formation games in general.
That is, we
propose a model where agents behave altruistically
to \emph{all their friends}, not only to the friends in 
the same coalition.  
Not restricting to hedonic games,
we aim
to capture %
a more natural
notion of altruism %
where
none of an agent's friends is %
excluded from her 
altruistic behavior.

\begin{example}
  \label{exa:ex_intro}
To become acquainted with this idea of altruism, consider the 
coalition formation game that is represented by the \emph{network of friends}
in Figure~\ref{fig:ex_intro}.
\begin{figure}[t] %
	\centering
	\begin{tikzpicture}
		\node (1) at (0,0) { $1$};
		\node (2) at (1,0) {$2$};
		\node (3) at (2,0) {$3$};
		\node (4) at (3,0) {$4$};
		\draw (1) -- (2);
		\draw (2) -- (3);
		\draw (3) -- (4);
	\end{tikzpicture}
	\caption{\label{fig:ex_intro} Network of friends for
          Example~\ref{exa:ex_intro}}
\end{figure}
For the coalition structures 
$\Gamma=\{\{1,2,3\},\{4\}\}$ and $\Delta=\{\{1,2,4\},\{3\}\}$,
it is clear that player $1$ is indifferent 
between coalitions $\{1,2,3\}$ and $\{1,2,4\}$ 
under friend-oriented preferences, as both coalitions contain
$1$'s only friend (player~$2$) and one of $1$'s enemies (either $3$ or~$4$).
Under altruistic hedonic preferences \cite{ngu-rey-rey-rot-sch:c:altruistic-hedonic-games},
however, player
$1$ behaves altruistically to her friend $2$ (who is friends with $3$ but not with $4$) and
therefore prefers $\{1,2,3\}$ to $\{1,2,4\}$.
Now, consider the slightly modified coalition structures 
$\Gamma'=\{\{1\},\{2,3\},\{4\}\}$ and $\Delta'=\{\{1\},\{2,4\},\{3\}\}$.
Intuitively, one would still expect $1$ to 
behave altruistically to her friend~$2$. 
However,
under any \emph{hedonic} preference 
(which requires the players' preferences to depend \emph{only} on their
own coalitions),
player $1$ (being in the same coalition for both $\Gamma'$ and~$\Delta'$)
must be indifferent between $\Gamma'$ and~$\Delta'$.
\end{example}

In order to model
altruism globally, 
we release the restriction to hedonic games
and 
introduce %
\emph{altruistic coalition formation games}
where agents behave altruistically to all their friends,
independently of 
their current coalition.

\subsection{Related Work}
Coalition formation games, as considered here, are closely related to the subclass of \emph{hedonic games} 
which has
been broadly studied in the
literature, addressing the issue of compactly representing preferences,
conducting axiomatic analyses, 
dealing with different notions of stability, 
and investigating
the computational complexity of the associated problems
(see, e.g., the book chapter by
Aziz and Savani~\cite{azi-sav:b:handbook-comsoc-hedonic-games}). 

Closest related to our work are the 
\emph{altruistic hedonic games} by Nguyen \emph{et al.}~\cite{ngu-rey-rey-rot-sch:c:altruistic-hedonic-games}
(see also the related minimization-based variant by
Wiechers and Rothe~\cite{wie-rot:c:stability-in-minimization-based-altruistic-hedonic-games}),
which we modify to obtain our more general models of altruism.
Based on the model due to Nguyen \emph{et al.}~\cite{ngu-rey-rey-rot-sch:c:altruistic-hedonic-games},
Schlueter and Goldsmith~\cite{sch-gol:c:super-altruistic-hedonic-games}
defined \emph{super altruistic hedonic games}
where
friends have a different impact on an agent based on their  
distances in the underlying network of friends.  
More recently,
Bullinger and Kober~\cite{bul-kob:c:loyalty-in-cardinal-hedonic-games}
introduced \emph{loyalty in cardinal hedonic games} where
agents are loyal to all agents in
their so-called loyalty set.
In their model, the utilities of the agents in the loyalty set are aggregated by taking the minimum.
They then study the loyal variants of common 
classes of cardinal hedonic games
such as additively separable %
and friend-oriented hedonic games.\footnote{Note that
their loyal variant of symmetric friend-oriented hedonic games is 
equivalent to the minimization-based altruistic hedonic games under equal treatment 
introduced by Wiechers and Rothe~\cite{wie-rot:c:stability-in-minimization-based-altruistic-hedonic-games}.}

Altruism has also been studied for \emph{non\-cooperative} games.  
Most prominently,
Ashlagi \emph{et al.}~\cite{ash-kry-ten:c:social-context-games}
introduced \emph{social context games} where a social context is applied to a strategic game
and 
the costs in the resulting game depend on
the original costs
and
a graph of neighborhood.  
Their so-called \emph{MinMax collaborations} (where players seek to minimize the maximal cost of 
their own and their neighbors)
are related to our minimization-based equal-treatment model.
Still, the model of Ashlagi \emph{et al.}~\cite{ash-kry-ten:c:social-context-games} 
differs from ours in that they consider
\emph{non}cooperative games.
Other work considering non\-cooperative games with social networks
is due to 
Bilò \emph{et al.}~\cite{bil-cel-fla-gal:j:social-context-congestion-games}
who study social context games
for other underlying strategic games than Ashlagi \emph{et al.}~\cite{ash-kry-ten:c:social-context-games},
Hoefer \emph{et al.}~\cite{hoe-pen-pol-sko-voe:c:considerate-equilibrium}
who study \emph{considerate equilibria} in strategic games, and
Anagnostopoulos \emph{et al.}~\cite{ana-bec-kei-sch:c:inefficiency-of-games-with-social-context}
who study \emph{altruism} and \emph{spite} in strategic games.
Further
work studying altruism in noncooperative games without social networks
is due to Hoefer and
Skopalik~\cite{hoe-sko:c:altruism-atomic-congestion-games},
Chen \emph{et
	al.}~\cite{che-kei-kem-sch:c:robust-price-of-anarchy-altruistic-games},
Apt and
Sch\"afer~\cite{apt-sch:j:selfishness-level-strategic-games},
and Rahn and
Sch\"afer~\cite{rah-sch:c:bounding-inefficiency-altruism-social-contribution-games}.

\subsection{Our Contribution}
Conceptually, we extend the models of altruism proposed by
Nguyen \emph{et al.}~\cite{ngu-rey-rey-rot-sch:c:altruistic-hedonic-games}
and Wiechers and Rothe~\cite{wie-rot:c:stability-in-minimization-based-altruistic-hedonic-games}
from hedonic games to general coalition formation games.  
We argue how this captures a more global notion of altruism 
and show that our models fulfill some desirable properties that
are violated by the previous models.
We then study the common stability concepts in this model and
analyze the associated verification and existence problems in terms of
their computational complexity.

This work %
extends a preliminary version that appeared in the 
proceedings of the \emph{29th International Joint Conference on Artificial 
Intelligence} (IJCAI'20)~\cite{ker-rot:c:altruism-in-coalition-formation-games}.
Parts of this work were also presented at 
the 16th and 17th International Symposium on Artificial Intelligence and Mathematics (ISAIM'20 and ISAIM'22) %
and at
the \emph{8th International Workshop on Computational Social Choice} (COMSOC'21),
each with nonarchival proceedings. 

\section{The Model}
\label{sec:model}
\OMIT{
\section{Preliminaries}

We will now provide some basic definitions. 
First, we will define \emph{coalition formation games}. 
After that, we will explain 
the \emph{``friends and enemies'' encoding} due to
Dimitrov \emph{et al.}~\cite{dim-bor-hen-sun:j:core-stability-hedonic-games} 
(see also the work of
Sung and Dimitrov~\cite{sun-dim:j:core-membership-hedonic-games}),
which we will use for the representation of the players' preferences.
} %
In \textit{coalition formation games}, players divide into groups based on their preferences. 
Before introducing altruism,
we now give some foundations of such games.

\subsection{Coalition Formation Games}
Let $N=\{1,\dots,n\}$ be a set of \emph{agents} (or \emph{players}). 
Each subset of $N$ is called a \emph{coalition}.
A \emph{coalition structure} $\Gamma$ is a partition of~$N$,
and we denote the set
of all possible coalition structures for
$N$ by~$\mathcal{C}_{N}$.  For a player $i\in N$ and a coalition
structure $\Gamma\in\mathcal{C}_{N}$, $\Gamma(i)$ denotes the unique coalition in
$\Gamma$ containing~$i$.
Now, a \emph{coalition formation game (CFG)}
is a pair $(N,\succeq)$, where $N=\{1,\dots,n\}$ is a set of agents,
${\succeq}={(\succeq_1,\dots,\succeq_n)}$
is a profile of preferences, and
every preference ${\succeq_i} \in \coalstr \times \coalstr$
is a complete weak order over
all possible coalition structures.
Given two coalition structures~$\Gamma,\ \Delta \in \mathcal{C}_N$, we say that $i$ \textit{weakly prefers} $\Gamma$ to $\Delta$ if $\Gamma \succeq_i \Delta$. When $\Gamma \succeq_i \Delta$ but not $\Delta \succeq_i \Gamma$, we say that $i$ \textit{prefers} $\Gamma$ to $\Delta$ (denoted by $\Gamma \succ_i \Delta$), and we say that $i$ is \textit{indifferent} between $\Gamma$ and $\Delta$ (denoted by $\Gamma \sim_i \Delta$) if $\Gamma \succeq_i \Delta$ and $\Delta \succeq_i \Gamma$. %

Note that
\emph{hedonic games} are a special case of coalition formation games where 
the agents' preference relations only depend on the coalitions containing
themselves.
In a hedonic game~$(N,\succeq)$, agent $i\in N$ is indifferent between any two coalition structures $\Gamma$ and $\Delta$ as long as her coalition is the same, i.e., 
$\Gamma(i)=\Delta(i) \Longrightarrow \Gamma \sim_i \Delta$.
Therefore, 
the preference order of any agent $i\in N$ in a hedonic game~$(N,\succeq)$ 
is usually represented by a complete weak order %
over %
the set of coalitions containing~$i$.

\subsection{The ``Friends and Enemies'' Encoding}
Since $\vert\coalstr\vert$, the number of all possible coalition structures,
is
extremely large in the number of agents,\footnote{
	The number of possible partitions of a 
	set with $n$ elements
	equals the $n$-th Bell number \cite{bel:j:the-iterated-exponential-integers,rot:j:the-number-of-partitions-of-a-set}, defined as
        $B_n = \sum_{k=0}^{n-1} \binom{n-1}{k} B_k$ with $B_0 = B_1 = 1$.
        For example, for $n=10$ agents, we have $B_{10} = 115,975$
        possible coalition structures.
	\label{foo:bell}} 
it is not
reasonable to ask every agent for her complete preference
over~$\coalstr$.
Instead, we are looking for a way to compactly represent the agents' preferences.
In the literature, many such representations
have been proposed %
for hedonic games,
such as
the \emph{additive encoding}
\cite{%
sun-dim:j:computational-complexity-additive-hedonic-games,azi-bra-see:j:computing-desirable-properties-in-hedonic-games,woe:j:hardness-for-core-stability-in-hedonic-games},
the \emph{singleton encoding} due to
Cechl{\'{a}}rov{\'{a}} and Romero-Medina~\cite{cec-rom:j:stability-coalition-formation-games}
and further studied by
Cechl{\'{a}}rov{\'{a}} and Hajdukov{\'{a}}~\cite{cec-haj:j:computational-complexity-stable-partitions-b-preferences},
the \emph{friends-and-enemies encoding} due to
Dimitrov \emph{et al.}~\cite{dim-bor-hen-sun:j:core-stability-hedonic-games},
and
FEN-hedonic games due to
Kerkmann
\emph{et al.}~\cite{ker-lan-rey-rot-sch-sch:j:hedonic-games-with-ordinal-preferences-and-thresholds}
and also used by
Rothe \emph{et al.}~\cite{rot-sch-sch:j:borda-induced-hedonic-games}.
Here, we
use the \emph{friends-and-enemies encoding} %
due to
Dimitrov \emph{et al.}~\cite{dim-bor-hen-sun:j:core-stability-hedonic-games}.
We focus on their friend-oriented model and will later adapt it to our
altruistic model.
 
In the friend-oriented model, the preferences of the agents in $N$ are
given by a network of friends, i.e., a (simple) undirected graph $G=(N,A)$
whose vertices are the players and where two players $i,j\in N$ are
connected by an edge $\{i,j\}\in A$ exactly if they are each other's
friends.  Agents
not connected by an edge consider each other as enemies.  For an agent
$i\in N$, we denote the set of $i$'s friends by $F_i = \{j\in
N \condition \{i,j\}\in A \}$ and the set of $i$'s enemies by $E_i
= N \setminus (F_i\cup\{i\})$.
Under %
\emph{friend-oriented preferences}
as defined by Dimitrov
\emph{et al.}~\cite{dim-bor-hen-sun:j:core-stability-hedonic-games}, 
between any two coalitions players prefer the coalition with more friends,
and if there are equally many friends in both coalitions,
they prefer the coalition with fewer enemies:
\[
C\succeq_i^F D \iff \vert C\cap F_i\vert > \vert D\cap F_i\vert \text{ or }
(\vert C\cap F_i\vert = \vert D\cap F_i\vert \text{ and } \vert C\cap E_i\vert \leq \vert D\cap E_i\vert).
\]

This can also be represented additively.
Assigning a value of $n$ to each friend 
and a value of $-1$ to each enemy,
agent $i\in N$ values %
coalition $C$ containing herself
with
$v_i(C) = n\vert C \cap F_i\vert - \vert C \cap E_i\vert$.
Note that 
$-(n-1)\leq v_i(C) \leq n
(n-1)$, and
$v_i(C) > 0$ if and only if there is at least one friend of $i$'s in~$C$.
For a given coalition structure $\Gamma\in \mathcal{C}_{N}$, we also write $v_i(\Gamma)$ for player $i$'s value of $\Gamma(i)$. %

Furthermore, we denote the sum of the values of $i$'s friends by
$\friendsum{i}{\Gamma} = \sum_{f \in F_i} v_f(\Gamma)$.
Analogously,
we also define
$\friendisum{i}{\Gamma} = \sum_{f \in F_i\cup \{i\}} v_f(\Gamma)$,
$\friendmin{i}{\Gamma} = \min_{f \in F_i} v_f(\Gamma)$, 
and $\friendimin{i}{\Gamma} = \min_{f \in F_i\cup \{i\}} v_f(\Gamma)$.

\subsection{Three Degrees of Altruism}

When we now define altruistic coalition formation games based on the
friend-oriented preference model, we consider the same three degrees
of altruism that
Nguyen \emph{et al.}~\cite{ngu-rey-rey-rot-sch:c:altruistic-hedonic-games}
introduced for altruistic hedonic games.
However,
we adapt them to our model, extending 
the agents' altruism to \emph{all} their friends, not only to their
friends in the same coalition.

\begin{itemize}
	\item \textbf{Selfish First (SF):}
	Agents first rank coalition structures based on their own
        valuations. Only in the case of a tie between two coalition
        structures, their friends' valuations are considered as well.
	\item \textbf{Equal Treatment (EQ):} 
	Agents treat themselves and their friends the same. 
	That means that an agent $i\in N$ and all of $i$'s friends 
	have the same impact on $i$'s utility for a coalition structure.
	\item \textbf{Altruistic Treatment (AL):}
	Agents first rank coalition structures based on their friends'
        valuations.  They only consider their own valuations in the
        case of a tie.
\end{itemize}

\noindent
We further distinguish between a sum-based and a min-based aggregation of
some agents' valuations.
Formally, 
for an agent $i\in N$ and a coalition structure
$\Gamma\in \coalstr$, we denote $i$'s sum-based utility for $\Gamma$ under
SF by $\utilitysfsum_{i}(\Gamma)$, under 
EQ by $\utilityeqsum_{i}(\Gamma)$, and under
AL by $\utilityalsum_{i}(\Gamma)$, and
her min-based utility for $\Gamma$ under
SF by $\utilitysfmin_{i}(\Gamma)$, under 
EQ by $\utilityeqmin_{i}(\Gamma)$, and under 
AL by $\utilityalmin_{i}(\Gamma)$.
For a constant $M \geq n^3$, they are defined as
\begin{align*}
	&\utilitysfsum_{i}(\Gamma)= M \cdot v_i(\Gamma) + \friendsum{i}{\Gamma};
	&&\utilitysfmin_{i}(\Gamma)= M \cdot v_i(\Gamma) + \friendmin{i}{\Gamma}; \\
	&\utilityeqsum_{i}(\Gamma)= \friendisum{i}{\Gamma};
	&&\utilityeqmin_{i}(\Gamma)= \friendimin{i}{\Gamma};\\
	&\utilityalsum_{i}(\Gamma)= v_i(\Gamma) + M \cdot \friendsum{i}{\Gamma};
	&&\utilityalmin_{i}(\Gamma)= v_i(\Gamma) + M \cdot \friendmin{i}{\Gamma}.
\end{align*}
In the case of $F_i=\emptyset$, we define the minimum of the empty set to be zero.

For any 
coalition structures $\Gamma,\Delta\in\coalstr$,
agent $i$'s sum-based SF 
preference is
then defined by
$\Gamma \succeqsf_{i} \Delta \iff \utilitysfsum_{i}(\Gamma) \geq \utilitysfsum_{i}(\Delta)$.
Her other altruistic preferences 
($\succeqequal_{i}$; $\succeqal_{i}$; $\succeqsfmin_i$; $\succeqequalmin_i$;
and $\succeqalmin_i$) 
are defined analogously,
using 
the respective utility functions.
The factor~$M$, which is used for the SF and 
AL models, ensures that an agent's utility is first
determined by the agent's own valuation in the SF model and
first determined by the friends' valuations in the AL model.
Similarly as
Nguyen \emph{et al.}~\cite{ngu-rey-rey-rot-sch:c:altruistic-hedonic-games} prove the
corresponding properties in hedonic games, we can show that
for $M \geq n^3$, $v_i(\Gamma) > v_i(\Delta)$ implies
$\Gamma \succsf_{i} \Delta$ and $\Gamma \succsfmin_{i} \Delta$, and
for $M \geq n^2$,
$\friendsum{i}{\Gamma}>\friendsum{i}{\Delta}$ implies
$\Gamma \succal_{i} \Delta$
while
$\friendmin{i}{\Gamma}>\friendmin{i}{\Delta}$ implies
$\Gamma \succalmin_{i} \Delta$.
An \emph{altruistic coalition formation game (ACFG)} is a coalition
formation game
where the agents' preferences were
obtained by a network of friends
via one of these
cases of altruism.
Hence, we distinguish between sum-based SF, sum-based EQ, sum-based AL, min-based SF, min-based EQ, and min-based AL ACFGs.  
For any ACFG, %
the players' utilities 
can obviously be computed in polynomial time.  

\section{Monotonicity and Other Properties in ACFGs}

Nguyen \emph{et al.}~\cite{ngu-rey-rey-rot-sch:c:altruistic-hedonic-games}
focus on altruism in hedonic games where an agent's utility only
depends on her own coalition.  
As we have already seen in Example~\ref{exa:ex_intro},
there are some aspects of altruistic behavior that cannot
be realized by hedonic games.
The following example shows that our model crucially differs
from the models due to
Nguyen \emph{et al.}~\cite{ngu-rey-rey-rot-sch:c:altruistic-hedonic-games}
and Wiechers and Rothe~\cite{wie-rot:c:stability-in-minimization-based-altruistic-hedonic-games}.

\begin{example} \label{ex:motivating2}
	Consider an ACFG $(N,\succeq)$ with
	the network of friends
	in Figure~\ref{fig:example_2} and
	\begin{figure}
		\centering
		\begin{tikzpicture}
			\node (5) at (0,0) {$5$};
			\node (1) at (1.1,0) {$1$};
			\node (2) at (2.2,0) {$2$};
			\node (3) at (3.3,0) {$3$};
			\node (4) at (3.3,0.8) {$4$};
			\node (6) at (0,0.8) {$6$};
			
			\node (7)  at (0.85,0.8) {$7$};
			\node (8)  at (1.35,0.8) {$8$};
			\node (9)  at (1.85,0.8) {$9$};
			\node (10) at (2.35,0.8) {$10$};
			
			\draw (1) -- (5);
			\draw (1) -- (6);
			\draw (1) -- (2);
			\draw (2) -- (3);
			\draw (3) -- (4);
			\draw (4) -- (2);
		\end{tikzpicture}
		\caption{\label{fig:example_2}
			Network of friends for %
			Example~\ref{ex:motivating2} }
	\end{figure}
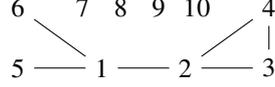
	the coalition structures
	$\Gamma=\{\{1,2\},\{3\},\{4\},\ldots,\{10\}\}$ and $\Delta=\{\{1,5,\ldots %
	,10\},\{2,3,4\}\}$. 
	We will now compare agent $1$'s preferences for these two coalition structures
	under our altruistic models
	to $1$'s preferences 
	under the altruistic hedonic models \cite{ngu-rey-rey-rot-sch:c:altruistic-hedonic-games,wie-rot:c:stability-in-minimization-based-altruistic-hedonic-games}.
	Table~\ref{tab:motivating2} shows 
	all relevant values that are needed to compute the utilities 
	of agent $1$. %
	\begin{table}
	  \caption{\label{tab:motivating2}
            Values for the game in Example~\ref{ex:motivating2}
            with the network of friends
	    in Figure~\ref{fig:example_2}}
		\centering
			\begin{tabular}{c|cccc|cc|cc}
				\toprule
				&$v_1$
				&$v_2$
				&$v_5$
				&$v_6$
				&$\mathrm{sum}^F_1$ & 
				$\mathrm{sum}^{F+}_1$ &
				$\mathrm{min}^F_1$ &
				$\mathrm{min}^{F+}_1$ \\
				\midrule
				$\Gamma$ 
				& $10$  &  $10$  &  $0$   & $0$
				&$10$& $20$ & $0$ & $0$ \\[1ex]
				$\Delta$ 
				& $16$  &  $20$  &  $5$   & $5$
				&$30$ & $46$ & $5$ & $5$ \\
				\bottomrule
			\end{tabular}
	\end{table}
	
	One can observe that agent $1$ and all her friends assign a greater value 
	to $\Delta$ than to~$\Gamma$. 
	Consequently, also the aggregations of the 
	friends' values 
	($\mathrm{sum}^F_1$, 
	$\mathrm{sum}^{F+}_1$,
	$\mathrm{min}^F_1$,
	$\mathrm{min}^{F+}_1$) are greater for~$\Delta$.
	Hence, %
	$1$ prefers $\Delta$ to $\Gamma$ 
	under all our sum-based and min-based altruistic preferences.
	
	The hedonic models due to Nguyen \emph{et al.}~\cite{ngu-rey-rey-rot-sch:c:altruistic-hedonic-games} and
	Wiechers and Rothe~\cite{wie-rot:c:stability-in-minimization-based-altruistic-hedonic-games}, however,
	are blind to the fact that 
	agent $1$ and all her friends 
	are better off in
	$\Delta$ than in~$\Gamma$.
	Under their altruistic hedonic preferences,
	player $1$ compares the two coalition structures $\Gamma$ and $\Delta$ only
	based on her own coalitions 
	$\Gamma(1)=\{1,2\}$ and $\Delta(1)=\{1,5,\ldots %
	,10\}$.
	She then only considers her friends that are in the same coalition,
	i.e., player $2$ for $\Gamma$ and players $5$ and $6$ for $\Delta$.
	This
	leads to 
	$1$ preferring $\Gamma(1)$ to $\Delta(1)$ under
	altruistic
	hedonic EQ and AL
	preferences.
	In particular, the average (and minimum) valuation of $1$'s friends in $\Gamma(1)$
	is $10$ while
	the average (and minimum) valuation of $1$'s friends in $\Delta(1)$
	is~$5$.	
	Also considering $1$'s own value for EQ,
	the average (and minimum) in $\Gamma(1)$ is $10$ while
	the average (respectively, minimum) value
        in $\Delta(1)$ is $8.\overline{6}$ (respectively,~$5$).
\end{example}

\subsection{Some Basic Properties}

As we have seen in Example~\ref{ex:motivating2}, 
altruistic \emph{hedonic} games~\cite{ngu-rey-rey-rot-sch:c:altruistic-hedonic-games,wie-rot:c:stability-in-minimization-based-altruistic-hedonic-games}
allow for
players that prefer coalition structures
that make themselves and all their friends worse off.
To avoid this kind of unreasonable behavior,
we focus on general coalition formation games.
In fact, all our altruistic \emph{coalition-formation} preferences 
fulfill unanimity:
For an ACFG $(N,\succeq)$ and a player $i \in N$, 
we say that $\succeq_i$ is \emph{unanimous} if,
for any two coalition structures $\Gamma,\Delta\in \coalstr$,
$v_a(\Gamma)>v_a(\Delta)$ for each $a\in F_i\cup
\{i\}$ implies $\Gamma\succ_i\Delta$.

This property crucially distinguishes our preference models 
from the corresponding altruistic \emph{hedonic}
preferences, which are not unanimous under EQ or AL preferences, as
Example~\ref{ex:motivating2} shows.
Note that
Nguyen \emph{et al.}~\cite{ngu-rey-rey-rot-sch:c:altruistic-hedonic-games}
define a restricted version of unanimity in altruistic hedonic games by
considering only the agents' own coalitions.
Other desirable properties that were studied by
Nguyen \emph{et al.}~\cite{ngu-rey-rey-rot-sch:c:altruistic-hedonic-games}
for altruistic hedonic preferences
can be generalized to
coalition formation games. 
We show that these desirable properties
also hold for our %
models.
First, we collect some basic observations:

\OMIT{
\section{Computing the Utilities}

For any ACFG %
(given by a network of
friends), the players' utilities can obviously be computed in polynomial time.  
We give the following proposition
which can be used to compute the utilities directly from the network of
friends.
The proof is again omitted due to
space constraints.

\begin{proposition}\label{net-calc}
	Let $G=(N,A)$ be a network of friends, $i\in N$ a player,
	and $\Gamma$ a coalition structure.
	Let further
	$\lambda$ be the number of friends that $i$ has in her
	coalition (i.e.,
	$\lambda=
	\bigl\vert\{\{i,j\}\in A \condition j\in \Gamma(i)\}\bigr\vert=
	\bigl\vert \Gamma(i) \cap F_i \bigr\vert$),
	$\mu$ be the number of edges between friends of $i$ that are together in a coalition (i.e.,
	$\mu= \vert\{\{j,k\}\in A \condition j\in F_i,\ k\in F_i,
	\ k\in \Gamma(j)\}\vert
	= \frac{1}{2} \sum_{j\in F_i} \vert F_i\cap F_j\cap\Gamma(j)\vert$),
	and
	$\nu$ be the number of edges between friends of $i$ and enemies of $i$ that are together in a coalition (i.e.,
	$\nu=\vert\{\{j,k\}\in A \condition j\in F_i,\ k\notin F_i,
	\ k\in \Gamma(j)\}\vert
	=\sum_{j\in F_i} \vert E_i\cap F_j\cap\Gamma(j)\vert$).
	Then, $i$'s utility can be computed by using 
	$v_i(\Gamma) = (n+1)\lambda-\vert\Gamma(i)\vert+1$ and
	$\friendsum{i}{\Gamma} =
	(n+1)(2\mu+\nu+\lambda)+\vert F_i\vert-\sum_{f\in F_i}\vert\Gamma(f)\vert$.
	
\end{proposition}
} %

\begin{observation}\label{obs:basic-properties}
	Consider any ACFG $(N,\succeq)$ with an underlying network of friends~$G$.
	\begin{enumerate}
		\item All preferences $\succeq_i$, $i\in N$, are reflexive and transitive.
		\item For any player $i\in N$ and any two coalition structures 
		$\Gamma,\Delta\in\coalstr$,
		it can be decided in polynomial time (in the number of agents) whether 
		$\Gamma \succeq_i \Delta$. \label{obs:pol}
		\item The preferences $\succeq_i$, $i\in N$, only depend on the 
		structure of~$G$.
	\end{enumerate}
\end{observation}

Note that the third statement of
Observation~\ref{obs:basic-properties} implies that the properties that 
Nguyen \emph{et al.}~\cite{ngu-rey-rey-rot-sch:c:altruistic-hedonic-games}
call \emph{anonymity} and \emph{symmetry} are both satisfied in ACFGs.
Another desirable property they consider is called
\emph{sovereignty of players} and inspired by the axiom of
\emph{``citizens' sovereignty''} from social choice theory:\footnote{Informally
  stated, a voting rule satisfies \emph{citizens' sovereignty} if every
  candidate can be made a winner of an election for a suitably chosen 
  preference profile.}
	Given a set of agents $N$, a coalition structure
	$\Gamma\in \mathcal{C}_{N}$, and 
	an agent~$i\in N$, we say that \emph{sovereignty of players} is
        satisfied if there is a network of friends $G$ on $N$ 
	such that $\Gamma$ is $i$'s
	most preferred coalition structure in any ACFG induced by~$G$. 

\begin{proposition}
  ACFGs satisfy sovereignty of players under all sum-based and min-based
  SF, EQ, and AL altruistic preferences.
\end{proposition}

\begin{proofs}
	Sovereignty of players in ACFGs can be shown with an analogous construction as in the proof of
	Nguyen \emph{et al.}~\cite[Theorem~5]{ngu-rey-rey-rot-sch:c:altruistic-hedonic-games}:
	For a given set of players~$N$, player $i\in N$, and coalition structure $\Gamma\in \coalstr$,
	we construct a network of friends where all players in $\Gamma(i)$ are friends of each other while there are no other friendship relations. Then $\Gamma$ is $i$'s 
	(nonunique) most preferred coalition structure
	under all sum-based and min-based SF, EQ, and AL altruistic preferences.~\end{proofs}

\subsection{Monotonicity}

The next property describes the monotonicity of preferences
and further distinguishes our models from altruistic hedonic games.
In fact,
Nguyen \emph{et al.}~\cite{ngu-rey-rey-rot-sch:c:altruistic-hedonic-games}
define two types of monotonicity, which we here adapt to our setting.

\begin{definition}
	Consider any ACFG $(N,\succeq)$, 
	agents $i,j\in N$ with $j\in E_i$, and
	coalition structures $\Gamma,\Delta \in \mathcal{C}_{N}$.
	Let further
	$\succeq^{\prime}_i$
	be the preference relation resulting from 
	$\succeq_i$
	when $j$ turns from being $i$'s enemy to being $i$'s friend
	(all else remaining %
	equal).
	We say that $\succeq_i$ is 
	\begin{itemize}
		\item\emph{type-I-monotonic} if 
		(1)~$\Gamma \succ_i \Delta$,
                $j\in \Gamma(i)\cap \Delta(i)$, and
                $v_j(\Gamma)\geq v_j(\Delta)$
		implies $\Gamma \succ^{\prime}_i \Delta$, and
		(2)~$\Gamma \sim_i \Delta$,
                $j\in \Gamma(i)\cap \Delta(i)$, and
                $v_j(\Gamma)\geq v_j(\Delta)$
		implies $\Gamma \succeq^{\prime}_i \Delta$;
                
		\item \emph{type-II-monotonic} if 
		(1)~$\Gamma \succ_i \Delta$,
                $j\in \Gamma(i)\setminus \Delta(i)$, and
		$v_j(\Gamma)\geq v_j(\Delta)$
		implies
		$\Gamma \succ^{\prime}_i \Delta$, and
		(2)~$\Gamma \sim_i \Delta$,
                $j\in \Gamma(i)\setminus \Delta(i)$, and
		$v_j(\Gamma)\geq v_j(\Delta)$
		implies $\Gamma \succeq^{\prime}_i \Delta$. 
	\end{itemize}
\end{definition}

\begin{theorem}
  \label{thm:monotonicity}
  Let $(N,\succeq)$ be an ACFG.
\begin{enumerate}
\item If $(N,\succeq)$ is sum-based,
  its preferences satisfy type-I- and type-II-monotonicity.

\item If $(N,\succeq)$ is min-based,
  its preferences satisfy type-II- but not type-I-monotonicity.
\end{enumerate}
\end{theorem}

\begin{proofs}
		Let $(N,\succeq)$ be an ACFG with an underlying network of friends $G=(N,H)$. Consider
		$i\in N$, $\Gamma,\Delta \in \mathcal{C}_{N}$, and $j\in E_i$
		and denote with $G'=(N,H\cup\{\{i,j\}\})$
		the network of friends resulting from $G$ when
		$j$ turns from being $i$'s enemy to being $i$'s friend 
		(all else being equal).
                Let $(N,\succeq^{\prime})$ be the ACFG induced by~$G'$.
		For any agent $a\in N$ and coalition structure
                $\Gamma\in \coalstr$, denote $a$'s value for $\Gamma$ in~$G'$
                by~$v_a'(\Gamma)$, $a$'s 
		preference relation in $(N,\succeq^{\prime})$
                by~$\succeq^{\prime}_a$,
		and $a$'s friends and enemies in $(N,\succeq^{\prime})$
                by $F'_a$ and~$E'_a$, respectively.
		That is, we have 
		$F'_i=F_i\cup\{j\}$, $E'_i=E_i\setminus\{j\}$, 
		$F'_j=F_j\cup\{i\}$, and $E'_j=E_j\setminus\{i\}$.
		Further, $v_i'$, $v_j'$, and %
		$\succeq^{\prime}_i$
		might differ from 
		$v_i$, $v_j$, and %
		$\succeq_i$,
		while
		the friends, enemies, and values of all other players stay the same, 
		i.e., $F_a'=F_a$, $E_a'=E_a$, and $v_a'=v_a$
		for all $a\in N \setminus\{i,j\}$.
		
		\paragraph{Type-I-monotonicity under sum-based preferences.}
		Let $j\in \Gamma(i)\cap \Delta(i)$ and $v_j(\Gamma)\geq  v_j(\Delta)$. 
		It then holds that
		\[
                v_i'(\Gamma)= n\vert\Gamma(i) \cap F_i'\vert - \vert\Gamma(i) \cap E_i'\vert= n\vert\Gamma(i) \cap F_i\vert+n - \vert\Gamma(i) \cap E_i\vert+1=v_i(\Gamma)+n+1.
                \]
		Equivalently,
		$v_i'(\Delta)=v_i(\Delta)+n+1$,
		$v_j'(\Gamma)=v_j(\Gamma)+n+1$,
		and
		$v_j'(\Delta)=v_j(\Delta)+n+1$.
		Furthermore,
		\begin{align}
			\friendsumprime{i}{\Gamma} &= \sum_{a\in F_i'} v'_a(\Gamma)
			= \sum_{a\in F_i\cup\{j\}} v'_a(\Gamma) %
			= \sum_{a\in  F_i} v_a(\Gamma) + v_j'(\Gamma) \nonumber \\
			&= \friendsum{i}{\Gamma} + v_j(\Gamma)+n+1
			\textnormal{\hspace{1mm} and } \label{eq:sum-A}\\
			\friendsumprime{i}{\Delta} 
			&= \friendsum{i}{\Delta} + v_j(\Delta)+n+1. \label{eq:sum-B}
		\end{align}
		
		\smallskip\textbf{(1) sumSF:}
		If $\Gamma\succsf_i \Delta$ then
		either
		(i)~$v_i(\Gamma)=v_i(\Delta)$ and
		$\friendsum{i}{\Gamma}>\friendsum{i}{\Delta}$,
		or
		(ii)~$v_i(\Gamma)>v_i(\Delta)$.
		
		In case~(i), 
		$v_i(\Gamma)=v_i(\Delta)$ implies $v_i'(\Gamma)=v_i'(\Delta)$. 
		Applying %
		$\friendsum{i}{\Gamma}>\friendsum{i}{\Delta}$ and 
		$v_j(\Gamma)\geq  v_j(\Delta)$
		to~(\ref{eq:sum-A}) and~(\ref{eq:sum-B}),
		we get 
		$\friendsumprime{i}{\Gamma}>\friendsumprime{i}{\Delta}$.
		This together with $v_i'(\Gamma)=v_i'(\Delta)$
		implies $\Gamma \succ^{\mathit{sumSF}\prime}_{i} \Delta$.
		
		In case~(ii), 
		$v_i(\Gamma)>v_i(\Delta)$ implies $v_i'(\Gamma)>v_i'(\Delta)$. Hence, $\Gamma \succ^{\mathit{sumSF}\prime}_{i} \Delta$.
		
		If $\Gamma\simsf_{i} \Delta$ then 
		$v_i(\Gamma)=v_i(\Delta)$
		and
		$\friendsum{i}{\Gamma}=\friendsum{i}{\Delta}$.
		$v_i(\Gamma)=v_i(\Delta)$ implies $v_i'(\Gamma)=v_i'(\Delta)$.
		Applying %
		$\friendsum{i}{\Gamma}=\friendsum{i}{\Delta}$ and 
		$v_j(\Gamma)\geq  v_j(\Delta)$
		to~(\ref{eq:sum-A}) and~(\ref{eq:sum-B}),
		we get 
		$\friendsumprime{i}{\Gamma}\geq\friendsumprime{i}{\Delta}$.
		This together with $v_i'(\Gamma)=v_i'(\Delta)$ 
		implies $\Gamma \succeq^{\mathit{sumSF}\prime}_{i} \Delta$.
		
		\smallskip\textbf{(2) sumEQ:}
		If $\Gamma\succequal_i \Delta$ then
		$\friendsum{i}{\Gamma}+ v_i(\Gamma) 
		>
		\friendsum{i}{\Delta}+v_i(\Delta)$. 
		Using \eqref{eq:sum-A}, \eqref{eq:sum-B}, 
		$v_i'(\Gamma)=v_i(\Gamma)+n+1$,
		$v_i'(\Delta)=v_i(\Delta)+n+1$, and 
		$v_j(\Gamma)\geq  v_j(\Delta)$,
		this implies 
		$\friendsumprime{i}{\Gamma}+ v_i'(\Gamma) 
		>
		\friendsumprime{i}{\Delta}+v_i'(\Delta)$. 
		Hence, 
		$\Gamma \succ^{\mathit{sumEQ}\prime}_i \Delta$.

		If $\Gamma\simequal_{i} \Delta$,
		using the same equations,
		$\Gamma \succeq^{\mathit{sumEQ}\prime}_i \Delta$
		is implied.
		
		\smallskip\textbf{(3) sumAL:}
		If $\Gamma\succal_i \Delta$ then 
		either
		(i)~$\friendsum{i}{\Gamma}=\friendsum{i}{\Delta}$ and
		$v_i(\Gamma)>v_i(\Delta)$,
		or
		(ii)~$\friendsum{i}{\Gamma}>\friendsum{i}{\Delta}$.
		
		In case~(i), $\friendsum{i}{\Gamma}=\friendsum{i}{\Delta}$
		together with~\eqref{eq:sum-A},~\eqref{eq:sum-B}, and
		$v_j(\Gamma)\geq  v_j(\Delta)$
		implies 
		$\friendsumprime{i}{\Gamma}\geq\friendsumprime{i}{\Delta}$.
		Further, $v_i(\Gamma)>v_i(\Delta)$ together with 
		$v_i'(\Gamma)=v_i(\Gamma)+n+1$ and
		$v_i'(\Delta)=v_i(\Delta)+n+1$
		implies 
		$v_i'(\Gamma)>v_i'(\Delta)$.
		Altogether, this implies
		$\Gamma \succ^{\mathit{sumAL}\prime}_i \Delta$.
		
		In case~(ii), $\friendsumprime{i}{\Gamma}
		> \friendsumprime{i}{\Delta}$
		is implied
		and $\Gamma \succ^{\mathit{sumAL}\prime}_i \Delta$ follows.
		
		If $\Gamma\simal_{i} \Delta$ then
		$\friendsum{i}{\Gamma}=\friendsum{i}{\Delta}$ and
		$v_i(\Gamma)=v_i(\Delta)$.
		Using the same equations as before,
		$\Gamma \succeq^{\mathit{sumAL}\prime}_i \Delta$
		is implied.
	        
		\paragraph{Type-II-monotonicity under sum-based and min-based preferences.}
		Let $j\in \Gamma(i)\setminus \Delta(i)$
		and $v_j(\Gamma)\geq v_j(\Delta)$.
		It follows that 
		$v_i'(\Gamma)= v_i(\Gamma)+n+1$,
		$v_i'(\Delta)=v_i(\Delta)$,
		$v_j'(\Gamma)= v_j(\Gamma)+n+1$, and
		$v_j'(\Delta)=v_j(\Delta)$.
		Furthermore,
		\begin{align}
			\friendsumprime{i}{\Gamma} 
			&= \friendsum{i}{\Gamma} + v_j(\Gamma)+n+1, \label{eq:sum-C}\\
			\friendsumprime{i}{\Delta} 
			&= \friendsum{i}{\Delta} + v_j(\Delta), \label{eq:sum-D}\\
			\friendminextprime{i}{\Gamma}& %
			=\min\bigl( \friendmin{i}{\Gamma},v_j(\Gamma)+n+1 \bigr),\\
			\friendminextprime{i}{\Delta}& %
			=\min\bigl( \friendmin{i}{\Delta},v_j(\Delta) \bigr),\\
			\friendiminextprime{i}{\Gamma}
			&=\min\bigl( \friendmin{i}{\Gamma},v_j(\Gamma)+n+1, v_i(\Gamma)+n+1 \bigr), 
			\textnormal{\hspace{1mm} and }\\
			\friendiminextprime{i}{\Delta}&
			=\min\bigl( \friendmin{i}{\Delta},v_j(\Delta), v_i(\Delta) \bigr).
		\end{align}
		
		\smallskip\textbf{(1) sumSF and minSF:}
		If $\Gamma\succeq^{\mathit{SF}}_{i}\Delta$ then
		$v_i(\Gamma)\geq v_i(\Delta)$.
		Hence, $v_i'(\Gamma)= v_i(\Gamma)+n+1\geq v_i(\Delta)+n+1> v_i(\Delta)=v_i'(\Delta)$, 
		which implies $\Gamma \succ^{\mathit{SF}\prime}_{i} \Delta$.

		\smallskip\textbf{(2) sumEQ:}
		If $\Gamma\succeqequal_{i}\Delta$ then
		$\friendsum{i}{\Gamma}+v_i(\Gamma)\geq \friendsum{i}{\Delta}+v_i(\Delta)$.
		Together with \eqref{eq:sum-C},~\eqref{eq:sum-D},
		and $v_j(\Gamma)\geq v_j(\Delta)$ this implies 
		$\friendsumprime{i}{\Gamma}+v_j'(\Gamma) > \friendsumprime{i}{\Delta}+v_j'(\Delta)$.
		Hence, 
		$\Gamma \succ^{\mathit{sumEQ}\prime}_i \Delta$.
		
		\smallskip\textbf{(3) sumAL:}
		If $\Gamma\succeqal_{i}\Delta$ then
		$\friendsum{i}{\Gamma}\geq \friendsum{i}{\Delta}$.
		Together with \eqref{eq:sum-C},~\eqref{eq:sum-D},
		and $v_j(\Gamma)\geq v_j(\Delta)$ this implies 
		$\friendsumprime{i}{\Gamma} > \friendsumprime{i}{\Delta}$, so
		$\Gamma \succ^{\mathit{sumAL}\prime}_i \Delta$.

		\smallskip\textbf{(4) minEQ:}
                First, assume that $\Gamma\succequalmin_{i}\Delta$.
		We then have
		$\min\bigl( \friendmin{i}{\Gamma}, v_i(\Gamma) \bigr)>
		\min\bigl( \friendmin{i}{\Delta}, v_i(\Delta) \bigr)$.
		It follows that $\Gamma\succ^{\mathit{minEQ}\prime}_{i}\Delta$ because
		\begin{align}
			&\friendiminextprime{i}{\Gamma}=\min\bigl( \friendmin{i}{\Gamma},v_j(\Gamma)+n+1, v_i(\Gamma)+n+1 \bigr)
			\label{eq:type2minEQ}\\
			&> \min\bigl( \friendmin{i}{\Delta},v_j(\Gamma), v_i(\Delta) \bigr)
			\geq  \min\bigl( \friendmin{i}{\Delta},v_j(\Delta), v_i(\Delta) \bigr)
			\nonumber%
			= \friendiminextprime{i}{\Delta}.
			\nonumber
		\end{align}
		
		Second, assume $\Gamma\simequalmin_{i}\Delta$.
		Then 
		$\min\bigl( \friendmin{i}{\Gamma}, v_i(\Gamma) \bigr)=
		\min\bigl( \friendmin{i}{\Delta}, v_i(\Delta) \bigr)$.
		Similarly as in~\eqref{eq:type2minEQ}, it follows that
		$\friendiminextprime{i}{\Gamma} \geq \friendiminextprime{i}{\Delta}$.
		Hence, $\Gamma\succeq^{\mathit{minEQ}\prime}_{i}\Delta$.

		\smallskip\textbf{(5) minAL:} First, assume $\Gamma\succalmin_{i}\Delta$.
		Then either (i)~$\friendmin{i}{\Gamma}>\friendmin{i}{\Delta}$, or
		(ii)~$\friendmin{i}{\Gamma}=\friendmin{i}{\Delta}$ and $v_i(\Gamma)> v_i(\Delta)$.

		In case of (i), we get $\Gamma\succ^{\mathit{minAL}\prime}_{i}\Delta$ because
		\begin{align}
			\friendminextprime{i}{\Gamma}&= \min\bigl( \friendmin{i}{\Gamma},v_j(\Gamma)+n+1 \bigr)
			\label{eq:type2minAL}
			\geq \min\bigl( \friendmin{i}{\Gamma},v_j(\Delta)+n+1 \bigr)\\
			& > \min\bigl( \friendmin{i}{\Delta},v_j(\Delta) \bigr)
			= \friendminextprime{i}{\Delta}.
			\nonumber
		\end{align}
		
		In case of (ii), similarly as in~\eqref{eq:type2minAL}, we get
		$\friendminextprime{i}{\Gamma}\geq \friendminextprime{i}{\Delta}$.
		Furthermore, $v_i(\Gamma)> v_i(\Delta)$ implies $v_i'(\Gamma)> v_i'(\Delta)$.
		Hence, $\Gamma\succ^{\mathit{minAL}\prime}_{i}\Delta$.
		
		Second, assume that $\Gamma\simalmin_{i}\Delta$.
		Then 
		$\friendmin{i}{\Gamma}=\friendmin{i}{\Delta}$ and $v_i(\Gamma)= v_i(\Delta)$.
		Similarly as in~\eqref{eq:type2minAL}, we get 
		$\friendminextprime{i}{\Gamma}\geq \friendminextprime{i}{\Delta}$.
		Furthermore, $v_i(\Gamma)= v_i(\Delta)$ implies $v_i'(\Gamma)> v_i'(\Delta)$.
		Hence, $\Gamma\succ^{\mathit{minAL}\prime}_{i}\Delta$.

		\paragraph{Type-I-monotonicity under min-based preferences.}
		To see that $\succeqsfmin$ is not type-I-monotonic, consider the 
		game $\mathcal{G}_1$
                with the network of friends in Figure~\ref{subfig:mbcfgs-monotonicity-g4}.
		\begin{figure}
			\centering
			\subfloat[Network of $\mathcal{G}_1$]{\label{subfig:mbcfgs-monotonicity-g4}
				\begin{tikzpicture}[scale=0.8, >=stealth, -, looseness =.7, line width=0.5pt,
					rund/.style={circle, draw=black, inner sep=0pt,  minimum size=15pt},
					leer/.style={inner sep=2pt}]
					\node (1) at (0,0) [leer] {$1$};
					\node (2) at (0,1) [leer]{$2$};
					\node (3) at (1,0) [leer] {$3$};
					\node (4) at (1,1) [leer] {$4$};
					\node (5) at (2,0) [leer]{$5$};
					\node (6) at (2,1) [leer] {$6$};
					\node (n) at (-0.5,0) [leer] {};
					\node (n) at (2.5,0) [leer] {};
					
					\draw (1) -- (3);
					\draw (1) -- (4);
					\draw (3) -- (4);
					\draw (3) -- (5);
					\draw (4) -- (5);
				\end{tikzpicture}
			}
			\subfloat[Network of $\mathcal{G}_1'$]{\label{subfig:mbcfgs-monotonicity-g4strich}
				\begin{tikzpicture}[scale=0.8, >=stealth, -, looseness =.7, line width=0.5pt,
					rund/.style={circle, draw=black, inner sep=0pt,  minimum size=15pt},
					leer/.style={inner sep=2pt}]
					\node (1) at (0,0) [leer] {$1$};
					\node (2) at (0,1) [leer]{$2$};
					\node (3) at (1,0) [leer] {$3$};
					\node (4) at (1,1) [leer] {$4$};
					\node (5) at (2,0) [leer]{$5$};
					\node (6) at (2,1) [leer] {$6$};
					\node (n) at (-0.5,0) [leer] {};
					\node (n) at (2.5,0) [leer] {};
					
					\draw (1) -- (2);
					\draw (1) -- (3);
					\draw (1) -- (4);
					\draw (3) -- (4);
					\draw (3) -- (5);
					\draw (4) -- (5);
				\end{tikzpicture}
			}
			\subfloat[Network of $\mathcal{G}_2$]{\label{subfig:mbcfgs-monotonicity-g5}
				\begin{tikzpicture}[scale=0.8, >=stealth, -, looseness =.7, line width=0.5pt,
					rund/.style={circle, draw=black, inner sep=0pt,  minimum size=15pt},
					leer/.style={inner sep=2pt}]
					\node (1) at (0,0) [leer] {$1$};
					\node (2) at (0,1) [leer]{$2$};
					\node (3) at (1,0) [leer] {$3$};
					\node (4) at (1,1) [leer] {$4$};
					\node (5) at (2,0) [leer]{$5$};
					\node (n) at (-0.5,0) [leer] {};
					\node (n) at (2.5,0) [leer] {};
					
					\draw (1) -- (3);
					\draw (1) -- (4);
					\draw (3) -- (4);
				\end{tikzpicture}
			}
			\subfloat[Network of $\mathcal{G}_2'$]{\label{subfig:mbcfgs-monotonicity-g5strich}
				\begin{tikzpicture}[scale=0.8, >=stealth, -, looseness =.7, line width=0.5pt,
					rund/.style={circle, draw=black, inner sep=0pt,  minimum size=15pt},
					leer/.style={inner sep=2pt}]
					\node (1) at (0,0) [leer] {$1$};
					\node (2) at (0,1) [leer]{$2$};
					\node (3) at (1,0) [leer] {$3$};
					\node (4) at (1,1) [leer] {$4$};
					\node (5) at (2,0) [leer]{$5$};
					\node (n) at (-0.5,0) [leer] {};
					\node (n) at (2.5,0) [leer] {};
					
					\draw (1) -- (2);
					\draw (1) -- (3);
					\draw (1) -- (4);
					\draw (3) -- (4);
				\end{tikzpicture}
			}
			\caption{\label{fig:mbcfgs-monotonicity}Networks of friends in the
				proof of Theorem~\ref{thm:monotonicity}}
		\end{figure}
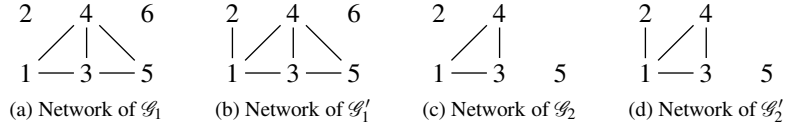
		Furthermore, consider the coalition structures 
		$\Gamma=\{\{1,2\},\{3,4,5\},\{6\}\}$ and
		$\Delta=\{\{1,2\},\{3,4,5,6\}\}$
		and players $i=1$ and $j=2$ with $2\in \Gamma(1)\cap \Delta(1)$, and $v_2(\Gamma)=-1= v_2(\Delta)$.
		It holds that 
		$v_1(\Gamma)=v_1(\Delta)=-1$,
		$\friendmin{1}{\Gamma}=2n$, and
		$\friendmin{1}{\Delta}=2n-1$.
		Hence,
		$\Gamma \succsfmin_1 \Delta$.
		
		Now, making $2$ a friend of $1$'s leads to the game 
		$\mathcal{G}_1'$ with the network of friends in Figure~\ref{subfig:mbcfgs-monotonicity-g4strich}.
		For this game, we have 
		$v_1'(\Gamma)=v_1'(\Delta)=n$ and
		$\friendminextprime{1}{\Gamma}=\friendminextprime{1}{\Delta}=n$.
		This implies 
		$\Gamma \sim^{\mathit{minSF\prime}}_1 \Delta$,
		which contradicts type-I-monotonicity.
		
		To see that $\succeqequalmin$ and $\succeqalmin$ are not type-I-monotonic, consider the 
		game $\mathcal{G}_2$ with the network of friends in Figure~\ref{subfig:mbcfgs-monotonicity-g5}.
		Consider the coalition structures 
		$\Gamma=\{\{1,2,3,4\},\{5\}\}$ and
		$\Delta=\{\{1,2,3,4,5\}\}$
		and players $i=1$ and $j=2$ with $2\in \Gamma(1)\cap \Delta(1)$, and $v_2(\Gamma)=-3>-4= v_2(\Delta)$.
		It holds that 
		$\friendimin{1}{\Gamma}=\friendmin{1}{\Gamma}=2n-1$, and
		$\friendimin{1}{\Delta}=\friendmin{1}{\Delta}=2n-2$.
		Hence,
		$\Gamma \succequalmin_1 \Delta$ and $\Gamma \succalmin_1 \Delta$.
		
		Now, making $2$ a friend of $1$'s leads to the game 
		$\mathcal{G}_2'$ with the network of friends in Figure~\ref{subfig:mbcfgs-monotonicity-g5strich}.
		For this game, we have 
		$\friendiminextprime{1}{\Gamma}=\friendminextprime{1}{\Gamma}=n$ and
		$\friendiminextprime{1}{\Delta}=\friendminextprime{1}{\Delta}=n$.
		This implies 
		$\Gamma \sim^{\mathit{minEQ\prime}}_1 \Delta$ and $\Gamma \sim^{\mathit{minAL\prime}}_1 \Delta$,
		contradicting type-I-monotonicity and
		completing the proof.~\end{proofs}

Note that
the hedonic models of altruism~\cite{ngu-rey-rey-rot-sch:c:altruistic-hedonic-games,wie-rot:c:stability-in-minimization-based-altruistic-hedonic-games} 
violate both type-I- and type-II-monotonicity 
for EQ and AL.
Hence,
it is quite
remarkable that all three degrees of our extended sum-based model of altruism
satisfy both types of monotonicity.

\section{Stability in ACFGs}

The main question in coalition formation games is which coalition
structures might form.  
There are several stability concepts that are well-studied for hedonic
games,
each indicating whether a given coalition structure would be accepted by
the agents or if there are other coalition structures that are more
likely to form. 
Although we consider more
general coalition formation games, we can easily adapt these definitions
to our framework.

Let $(N,\succeq)$ be an
ACFG with %
preferences ${\succeq} =(\succeq_1,\dots,\allowbreak\succeq_n)$ 
obtained from a network of friends via
one of the three degrees of altruism and with either sum-based or
min-based aggregation of the agents' valuations.
We use the following notation.
For a coalition structure~$\Gamma\in \coalstr$, a player $i \in N$, and a coalition $C\in\Gamma \cup \{\emptyset\}$, 
$\Gamma_{i\rightarrow C}$ denotes the coalition structure
that arises from $\Gamma$ when moving $i$ to~$C$, i.e.,
\[
\Gamma_{i\rightarrow C}=
\Gamma\setminus\{ \Gamma(i),C\}\cup\{\Gamma(i)\setminus\{i\},C\cup\{i\} \}.
\]

In addition, we use $\Gamma_{C\rightarrow\emptyset}$, with $C\subseteq N$, 
to denote the coalition structure that arises from $\Gamma$ when all players in $C$ leave their respective coalition and form a new one, i.e.,
\[
\Gamma_{C\rightarrow\emptyset}=\Gamma \setminus \{\Gamma(j)\condition j\in C\} \cup \{\Gamma(j)\setminus C\condition j\in C \} \cup \{C\}.
\]

Finally, for any two coalition structures $\Gamma,
\Delta\in \coalstr$, let $\#_{\Gamma \succ \Delta}=\vert\{i\in N\condition
\Gamma \succ_i \Delta\}\vert$ be the number of players that prefer
$\Gamma$ to~$\Delta$.
Now, we are ready to define the common stability notions.

\begin{definition}
  \label{def:stability-notions}
A coalition structure $\Gamma$ is said to be
\begin{itemize}
	\item \emph{Nash stable} if no player prefers moving to
	another coalition:
	\[
	(\forall i\in N)(\forall C\in \Gamma\cup \{\emptyset\})
	[ \Gamma \succeq_i \Gamma_{i\rightarrow C} ];
	\]
	\item \emph{individually rational} if 
	no player would prefer being alone:
	\[
	(\forall i\in N)[ \Gamma \succeq_i 
	\Gamma_{i\rightarrow \emptyset}
	];
	\]
	\item \emph{individually stable} if 
	no player prefers moving to another coalition
	and could deviate to it without harming any player in that coalition:
	\[
	(\forall i\in N)(\forall C\in \Gamma\cup\{\emptyset\})
	\bigl[ \Gamma\succeq_i \Gamma_{i\rightarrow C}
	\vee (\exists j\in C)[\Gamma \succ_j\Gamma_{i\rightarrow C} ] \bigr];
	\]
	\item \emph{contractually individually stable}
	if no player prefers another coalition 
	and could deviate to it without harming any
	player in the new or the old coalition:
	\[
        \hspace*{-2mm}
	(\forall i\in N)(\forall C\in \Gamma\cup\{\emptyset\})
	\bigl[ \Gamma\succeq_i \Gamma_{i\rightarrow C}
	\vee (\exists j\in C)[\Gamma \succ_j\Gamma_{i\rightarrow C} ]
	\vee (\exists k\in \Gamma(i))[\Gamma \succ_k\Gamma_{i\rightarrow C} ]
	\bigr];
	\]
	\item \emph{totally individually stable}
	if no player prefers another coalition 
	and could deviate to it without harming any
	other player:
	\[
	(\forall i\in N)(\forall C\in \Gamma\cup\{\emptyset\})
	\bigl[ \Gamma\succeq_i \Gamma_{i\rightarrow C}
	\vee (\exists l\in N\setminus\{i\})[\Gamma \succ_l\Gamma_{i\rightarrow C} ]
	\bigr];
	\]
	\item \emph{core stable} if
	no nonempty 
	coalition blocks~$\Gamma$:
	\[
	(\forall C\subseteq N, C\neq \emptyset)(\exists i\in C)[
	\Gamma \succeq_i \Gamma_{C\rightarrow\emptyset}];
	\]
	\item \emph{strictly core stable} if
	no coalition weakly blocks~$\Gamma$:
	\[
	(\forall C\subseteq N
	)
	(\exists i\in C)[
	\Gamma \succ_i \Gamma_{C\rightarrow\emptyset}]
	\vee
	(\forall i\in C)[
	\Gamma \sim_i \Gamma_{C\rightarrow\emptyset}];
	\]
	\item \emph{popular} if for every other coalition
	structure~$\Delta$, at least as many players prefer $\Gamma$
	to $\Delta$ as there are players who prefer $\Delta$
	to~$\Gamma$:
	\[
	(\forall \Delta \in \coalstr, \Delta  \neq \Gamma) \bigl[ \#_{\Gamma \succ \Delta} \geq
	\#_{\Delta \succ \Gamma} \bigr];
	\]
	\item \emph{strictly popular} if for every other coalition
	structure~$\Delta$, more players prefer $\Gamma$ to $\Delta$
	than there are players who prefer $\Delta$ to~$\Gamma$:
	\[
	(\forall \Delta \in \coalstr, \Delta  \neq \Gamma) \bigl[ \#_{\Gamma \succ \Delta} >
	\#_{\Delta \succ \Gamma} \bigr];
	\]
	\item \emph{perfect} if no player prefers
	any coalition structure to~$\Gamma$:
	\[
	(\forall i\in N)(\forall \Delta\in \mathcal{C}_{N})[\Gamma \succeq_i \Delta].
	\]
\end{itemize}
\end{definition}

\noindent
Note that ``totally individual stability'' is a new notion which we
introduce here. It
strengthens the notion of
contractually individual stability and makes sense in the context of
coalition formation games because players' preferences may also be
influenced by coalitions they are not part of.

We now study the associated \emph{verification} and \emph{existence
  problems}
in terms of their computational complexity.
We assume the reader to be familiar with the complexity classes $\pol$
(deterministic polynomial time), $\np$ (nondeterministic polynomial time)
and $\co\np$ (the class of complements of $\np$ sets).
For more background on computational complexity, we refer to, e.g.,
the textbooks by Garey and Johnson~\cite{gar-joh:b:int} and
Rothe~\cite{rot:b:cryptocomplexity}.
Given a stability concept $\alpha$, we define: 
\OMIT{
\EP{$\alpha$-Verification}
{An ACFG $(N,\succeq)$ and a coalition structure $\Gamma\in\coalstr$.}
{Does $\Gamma$ satisfy~$\alpha$?}

\EP{$\alpha$-Existence}
{An ACFG $(N,\succeq)$.}
{Does there exist a coalition structure $\Gamma\in\coalstr$ that
	satisfies~$\alpha$?}
} %
\begin{itemize}
\item \textsc{$\alpha$-Verification}: Given an ACFG $(N,\succeq)$ and a
coalition structure $\Gamma\in\coalstr$, does $\Gamma$ satisfy~$\alpha$?
\item \textsc{$\alpha$-Existence}: Given an ACFG $(N,\succeq)$, does there
exist a coalition structure $\Gamma\in\coalstr$ that satisfies~$\alpha$?
\end{itemize}

\OOMIT{
	\begin{table*}[t]
		\centering
		\begin{tabular}{lll}
			\hline \hline 
			$\alpha$ & {\rmfamily \textsc{Verification}} & {\rmfamily \textsc{Existence}} \\ \hline\midrule %
			individual rationality&   
			in \pol  & in \pol\ (trivially: YES) \\%
			Nash stability&     
			in \pol  & in \pol\ (trivially: YES) \\%
			(contractually/ totally) individual stability&     
			in \pol  & in \pol\ (trivially: YES) \\ \hline%
			core stability&  
			in \conp  & in \pol\ for SF (trivially: YES) \\%
			strict core stability& 
			in \conp  & in \pol\ for SF (trivially: YES)\\ \hline%
			popularity& 
			not trivial, in \conp  & not trivial \\%
			strict popularity&
			not trivial, in \conp, \conp-complete for SF  & not trivial \\ \hline%
			perfectness&        
			not trivial, in \conp  & not trivial, in \conp\ for EQ \\%
			& in \pol\ for SF & in \pol\ for SF \\%
			\hline\hline
		\end{tabular}
		\caption{Results for the Verification and Existence Problem}\label{tab:results}
	\end{table*}
} %

Table~\ref{tab:results_SF} summarizes the results for these
problems under sum-based and min-based SF preferences.
We will also give results for
EQ and AL
in this section.  
In Table~\ref{tab:results_SF}, however,
we only mark if the results for
EQ and AL
match those for
SF.

\begin{table}[t]
  \caption{Complexity results
    in sum-based and min-based SF ACFGs
	}\label{tab:results_SF}
	\centering
	\begin{threeparttable}[t]
	\centering
	\begin{tabular}{l@{\quad}l@{\quad}l}
		\toprule
		Stability notion $\alpha$ & {\rmfamily \textsc{$\alpha$-Verification}} & {\rmfamily \textsc{$\alpha$-Existence}} \\ \midrule %
		Individual rationality&
		in \pol \tnote{1} & trivial\tnote{1} \\%
		Nash stability&     
		in \pol\tnote{1}  & trivial\tnote{1} \\%
		Individual stability& 
		in \pol\tnote{1}  & trivial\tnote{1} \\ \midrule%
		Core stability&  
		\conp-complete\tnote{2}  & trivial \\%
		Strict 
		core stability& 
		in \conp\tnote{2}  & trivial\\ \midrule%
		Popularity& 
		\conp-complete\tnote{2}  & not trivial\tnote{1} \\%
		Strict 
		popularity&
		\conp-complete\tnote{2}  & \conp-hard \\ \midrule%
		Perfectness&        
		in \pol\tnote{2} & in \pol\tnote{3} \\%
		\bottomrule
	\end{tabular}
	\begin{tablenotes}
	\item[1]
          also holds for sum-based and min-based EQ and AL ACFGs
	\item[2]
          is in \conp\ for any ACFG
	\item[3]
          is in \conp\ for sum-based EQ ACFGs
	\end{tablenotes}
\end{threeparttable}
\end{table}

\subsection{Individual Rationality}
\label{subsec:IR}

Verifying individual rationality is easy: We just need to iterate
over all agents and compare two coalition structures in each
iteration. Since players' utilities can be computed in polynomial
time, individual rationality can be verified in time
polynomial in the number of agents.
The existence problem is trivial, since
$\Gamma=\{\{1\},\ldots,\{n\}\}$ is always individually rational.
Furthermore, we give the following characterization.

\begin{theorem}
	\label{char:IR}
	Given an ACFG $(N,\succeq)$, a coalition structure $\Gamma\in\mathcal{C}_N$ 
	is individually rational
	\begin{enumerate}
		\item under sum-based 
		SF, sum-based EQ, sum-based AL,
		min-based SF, or min-based AL preferences
		if and only if it holds for all players $i\in N$ that
		$\Gamma(i)$ contains a friend of $i$'s or $i$ is alone, formally:
		$(\forall i \in N)[\Gamma(i)\cap F_i\neq \emptyset \vee \Gamma(i)=\{i\}]$;
		\item under min-based EQ preferences
		if and only if
		for all players $i\in N$,
		$\Gamma(i)$ contains a friend of $i$'s or $i$ is alone or 
		there is a friend of $i$'s whose valuation of $\Gamma$
                is less than or equal to $i$'s valuation of $\Gamma$,
		formally:
		$(\forall i\in N) [%
		\Gamma(i)\cap F_i\neq \emptyset \vee \Gamma(i)=\{i\}
		\lor
		(\exists j\in F_i) [v_j(\Gamma)\leq v_i(\Gamma)]]$.
	\end{enumerate}
\end{theorem}

\begin{proofs}
	1. To show the implication from left to right, if $\Gamma$ is
        individually rational, we assume for the sake of contradiction
	that $\Gamma(i) \cap F_i = \emptyset$ and $\Gamma(i) \neq \{i\}$
        for some player $i\in N$. 
	First, we observe that for all $j\in F_i$ we have $v_j(\Gamma) = v_j(\Gamma_{i\rightarrow\emptyset})$, as their respective coalition is not affected by $i$'s move. 
	It directly follows that, for all considered models of altruism, 
	player $i$'s utilities for $\Gamma$ and $\Gamma_{i\rightarrow\emptyset}$ only 
	depend on her own valuation, which is
	greater for $\Gamma_{i\rightarrow\emptyset}$ than for $\Gamma$ (since there are enemies in $\Gamma(i)$ but not in $\Gamma_{i\rightarrow\emptyset}(i)$).
	Hence, $i$ prefers $\Gamma_{i\rightarrow\emptyset}$ to $\Gamma$, so $\Gamma$ is not individually rational. This is a contradiction. 

        The implication from right to left is obvious for all considered
        models of altruism.

        \smallskip
        2.
        From left to right, we have that $\Gamma$ is individually rational and, 
	for the sake of contradiction, we assume	
	that there is a player $i\in N$ with $\Gamma(i) \cap F_i = \emptyset$ and $\Gamma(i) \neq \{i\}$ %
	and for all $j\in F_i$ we have $v_j(\Gamma) > v_i(\Gamma)$.
        Since $i$ is the least satisfied player 
	in $F_i\cup\{i\}$,
	we have $\utilityeqmin_i(\Gamma) = v_i(\Gamma)$.
	With  $v_j(\Gamma_{i\rightarrow\emptyset}) = v_j(\Gamma)> v_i(\Gamma)$ for all $j\in F_i$ 
	and $v_i(\Gamma_{i\rightarrow\emptyset}) = 0 > v_i(\Gamma)$, we immediately obtain 
	$\utilityeqmin_i(\Gamma_{i\rightarrow\emptyset})> \utilityeqmin_i(\Gamma)$ and
	$\Gamma_{i\rightarrow\emptyset} \succequalmin_i {\Gamma}$. This is a contradiction to $\Gamma$ being individually rational.
        
	From right to left, we have to consider two cases. First, if $\Gamma(i)\cap F_i\neq \emptyset$ or $\Gamma(i)=\{i\}$ for some $i\in N$, we obviously have $\Gamma \succeqequalmin_i \Gamma_{i\rightarrow\emptyset}$. 
	Second, if $\Gamma(i) \cap F_i = \emptyset$ and $\Gamma(i) \neq \{i\}$, we know that
	there is at least one $j\in F_i$ with $v_j(\Gamma) \leq v_i(\Gamma)<0$. 
	Let $j'$ denote a least satisfied friend of $i$'s in $\Gamma$ (pick one randomly if there are more than one). 
	Since $\Gamma(i) \cap F_i = \emptyset$, %
	it holds that $\Gamma(j)=\Gamma_{i\rightarrow\emptyset}(j)$ for all $j\in F_i$. 
	Consequently, $j'$ is $i$'s least satisfied friend in both coalition structures and we have $\utilityeqmin_i(\Gamma) = v_{j'}(\Gamma) = v_{j'}(\Gamma_{i\rightarrow\emptyset}) = \utilityeqmin_i(\Gamma_{i\rightarrow\emptyset})$.
        Hence, $\Gamma \simequalmin_i \Gamma_{i\rightarrow\emptyset}$, so
	$\Gamma$ is individually~rational.~\end{proofs}

\subsection{Nash Stability}
\label{subsec:NS}

Since there are at most $\vert N\vert$ coalitions in a coalition structure $\Gamma\in  \coalstr$, we can verify
Nash stability
in polynomial time: We just iterate over all agents $i\in N$ and
all the (at most $\vert N\vert+1$) coalitions
$C\in \Gamma\cup \{\emptyset\}$ and
check whether $\Gamma \succeq_i \Gamma_{i\rightarrow C}$.  Since we
can check a player's altruistic preferences over any two coalition structures in
polynomial time and since we have at most a quadratic number of
iterations ($\vert N\vert\cdot (\vert N\vert+1)$), Nash
stability verification is in \pol\ for any ACFG.

Nash stability existence is trivially in \pol\ for any ACFG; indeed, the same example
that %
Nguyen \emph{et al.}~\cite{ngu-rey-rey-rot-sch:c:altruistic-hedonic-games} 
gave for
altruistic hedonic games works here as well.
Specifically, for $C=\{i \in N \condition F_i= \emptyset\}=\{c_1,\ldots,c_k\}$
the coalition structure $\{\{c_1\},\ldots,\{c_k\},N\setminus C\}$
is Nash stable. 

\OMIT{
\begin{theorem}
	\label{char:NS}
	Let $(N,\succeq)$ be an ACFG
	where the preferences were obtained by a network of friends via one of the three degrees of altruism. 
	A coalition structure $\Gamma$ is Nash stable
	if and only if 
$(\forall i\in N)(\forall C\in \Gamma\cup \{\emptyset\})
	\Bigl[ \vert \Gamma(i) \cap F_i\vert > \vert C\cap F_i\vert
        \vee
	\bigl( \vert \Gamma(i) \cap F_i\vert = \vert C\cap F_i\vert \wedge \vert\Gamma(i) \cap E_i\vert \leq \vert C\cap E_i\vert \bigr)
	\Bigr]$.
\end{theorem}
The proof of Theorem~\ref{char:NS} 
is also omitted here.
}

\subsection{Individual Stability}
\label{subsec:IS}

For individual stability, 
contractually individual stability, 
and
totally individual stability, existence is also trivially in \pol. Nash
stability implies all these three concepts, hence, the Nash stable
coalition structure given above is also (contractually; totally)
individually stable.

Verification is also in \pol\ for these stability concepts.  Similarly to Nash
stability, we can iterate over all players and all coalitions and check the
respective conditions in polynomial time.

\subsection{Core Stability and Strict Core Stability}
\label{subsec:CS}

We now turn to core stability and state some results for 
sum-based and min-based SF ACFGs.
We first show that (strict) core stability existence 
is trivial for SF ACFGs.

\begin{theorem}
	\label{thm:core-existence}
	Let $(N,\succeq^{SF})$ be a
	(sum-based or min-based) SF
	ACFG with the underlying network of friends $G$.  
	Let further $C_1,\ldots,
	C_k$ be the vertex sets of the connected components of $G$.  Then
	$\Gamma=\{C_1,\ldots, C_k\}$ is strictly core stable (and thus core
	stable).
\end{theorem}
\begin{proofs}
	For the sake of contradiction, assume that $\Gamma$ were not strictly core stable, i.e., that there is a coalition $D\neq \emptyset$ that weakly blocks $\Gamma$.
	Consider some player $i\in D$.
	Since 
	$i$
	weakly prefers deviating
	from $\Gamma(i)$ to~$D$,
	there have to be at least as many friends of $i$'s in $D$ as
	in~$\Gamma(i)$.
	Since $\Gamma(i)$ contains all of $i$'s friends, 
	$D$ also has to contain all friends of~$i$'s.
	Then all these friends of~$i$'s also have all their friends in $D$ for the same reason, and so on. Consequently, $D$ contains all players from the connected component $\Gamma(i)$, i.e., $\Gamma(i)\subseteq D$.
	
	Since $D$ weakly blocks $\Gamma$, $D$ cannot be equal to $\Gamma(i)$ and thus needs to contain some $\ell\notin \Gamma(i)$.
	Yet, this is a contradiction, as $\ell$ is an enemy of $i$'s and $i$ would prefer 
	$\Gamma$ to $\Gamma_{D\rightarrow\emptyset}$
	if $D$ contains the same number of friends as $\Gamma(i)$ but more enemies than~$\Gamma(i)$.~\end{proofs}

However, 
the coalition structure from %
Theorem~\ref{thm:core-existence} is not necessarily core stable under
EQ and AL preferences.

\begin{example}
	\label{ex:blocking-coalition}
	Let $N=\{1,\ldots,10\}$ and consider the
	network of friends $G$ shown in Figure~\ref{fig:blocking-coalition}.
	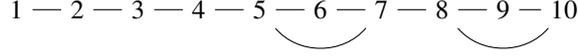
\begin{figure}[t]
		\centering
			\begin{tikzpicture}
				\node (1) at (0,0) {$1$};
				\node (2) at (0.8,0) {$2$};
				\node (3) at (1.6,0) {$3$};
				\node (4) at (2.4,0) {$4$};
				\node (5) at (3.2,0) {$5$};
				\node (6) at (4,0) {$6$};
				\node (7) at (4.8,0) {$7$};
				\node (8) at (5.6,0) {$8$};
				\node (9) at (6.4,0) {$9$};
				\node (10) at (7.2,0) {$10$};
				
				\draw (1) -- (2);
				\draw (2) -- (3);
				\draw (3) -- (4);
				\draw (4) -- (5);
				\draw (5) -- (6);
				\draw (6) -- (7);
				\draw (7) -- (8);
				\draw (8) -- (9);
				\draw (9) -- (10);
				\draw (10) to[out=-130,in=-50] (8);
				\draw (7) to[out=-130,in=-50] (5);
			\end{tikzpicture}
		\caption{\label{fig:blocking-coalition}Networks of friends for Example~\ref{ex:blocking-coalition}}
	\end{figure} 
	Consider the coalition structure consisting of the connected
        component of~$G$ (i.e., of only the grand coalition: $\Gamma=\{N\}$)
	and the coalition $C=\{8,9,10\}$.
	$C$ blocks $\Gamma$ under sum-based and min-based EQ and AL preferences.
	To see this, consider how players $7$, $8$, $9$, and $10$ value
	$\Gamma$ and $\Gamma_{C\rightarrow\emptyset}$: 
\begin{alignat*}{2}
	v_7(\Gamma) &= v_8(\Gamma) = 30-6 = 24,\qquad & 
	v_7(\Gamma_{C\rightarrow\emptyset}) &= 20-4 = 16, \\
	v_9(\Gamma) &= v_{10}(\Gamma) = 20-7 = 13,\qquad & 
	v_8(\Gamma_{C\rightarrow\emptyset}) &= v_9(\Gamma_{C\rightarrow\emptyset}) = v_{10}(\Gamma_{C\rightarrow\emptyset}) = 20.
\end{alignat*}
	We then obtain
        \begin{itemize}
	  \item $\friendisum{8}{\Gamma}=74<76=\friendisum{8}{\Gamma_{C\rightarrow\emptyset}}$ and
	$\friendisum{9}{\Gamma}=\friendisum{10}{\Gamma}=50<60=\friendisum{9}{\Gamma_{C\rightarrow\emptyset}}=\friendisum{10}{\Gamma_{C\rightarrow\emptyset}}$,
	so $\Gamma_{C\rightarrow\emptyset}\succequal_i \Gamma$ for all $i\in C$;
	  \item $\friendsum{8}{\Gamma}=50<56=\friendsum{8}{\Gamma_{C\rightarrow\emptyset}}$ and
	$\friendsum{9}{\Gamma}=\friendsum{10}{\Gamma}=37<40=\friendsum{9}{\Gamma_{C\rightarrow\emptyset}}=\friendsum{10}{\Gamma_{C\rightarrow\emptyset}}$,
	so $\Gamma_{C\rightarrow\emptyset}\succal_i \Gamma$ for all $i\in C$;
	  \item $\friendimin{8}{\Gamma}=\friendmin{8}{\Gamma}=13<16=\friendimin{8}{\Gamma_{C\rightarrow\emptyset}}=\friendmin{8}{\Gamma_{C\rightarrow\emptyset}}$ and 
	$\friendimin{9}{\Gamma}=\friendmin{9}{\Gamma}=\friendimin{10}{\Gamma}=\friendmin{10}{\Gamma}=13<20=\friendimin{9}{\Gamma_{C\rightarrow\emptyset}}=\friendmin{9}{\Gamma_{C\rightarrow\emptyset}}=\friendimin{10}{\Gamma_{C\rightarrow\emptyset}}=\friendmin{10}{\Gamma_{C\rightarrow\emptyset}}$,
	which implies $\Gamma_{C\rightarrow\emptyset}\succequalmin_i \Gamma$ and $\Gamma_{C\rightarrow\emptyset}\succalmin_i \Gamma$ for all $i\in C$.
        \end{itemize}
	Thus $C$ blocks $\Gamma$ 
	under sum-based and min-based EQ and AL preferences.
\end{example}

Turning to (strict) core stability verification,
we can show that this problem is hard under SF preferences, and 
we suspect that this hardness also extends to EQ and~AL.

\begin{theorem}\label{thm:core-ver-in-conp}
  Strict core stability verification and core stability verification
  are in \conp\ for any ACFG.  For (sum-based and min-based) SF ACFGs,
  core stability verification is even \conp-complete.
\end{theorem}

\begin{proofs}
  To see that strict core stability verification and
  core stability verification are in \conp,
	consider any
	coalition structure $\Gamma\in \coalstr$ in an
	ACFG~$(N,\succeq)$.
	$\Gamma$ is not (strictly) core stable if there is a coalition $C\subseteq N$ that (weakly) blocks~$\Gamma$.
	Hence, we nondeterministically guess a coalition $C\subseteq N$ and check whether $C$ (weakly) blocks~$\Gamma$.
	This can be done in polynomial time
	since the preferences of the agents in $C$ for the coalition structures $\Gamma$ and
	$\Gamma_{C\rightarrow\emptyset}$ can be verified in polynomial time for all our altruistic models.

To show \conp-hardness of core stability verification under min-based SF ACFGs, 
we use \rxthreec, which is a restricted variant of \xthreeclong\ and
known to be \np-complete \cite{gar-joh:b:int,gon:j:exact-cover-constrained}.
We provide a polynomial-time many-one reduction from \rxthreec\ to the complement of 
our verification problem.
Let $(B,\mathscr{S})$ be an instance of \rxthreec,
consisting of a set $B=\{1,\dots,3k\}$ and 
a collection $\mathscr{S}=\{S_1,\dots, S_{3k}\}$ of 3-element subsets of $B$,
where each element of $B$ occurs in exactly three sets in $\mathscr{S}$.
The question %
is  whether there exists 
an exact cover for~$B$ in~$\mathcal{S}$, i.e., 
a subset $\mathcal{S}'\subseteq\mathcal{S}$ with
$\vert\mathcal{S}'\vert = k$ and $\bigcup_{{S}\in{\mathcal{S}'}} S=B$.
We assume that $k>4$.

From $(B,\mathscr{S})$ we now construct the following ACFG.
The set of players is
	$N= \{\beta_b\condition b\in B\} \cup
	    \{\zeta_S, \alpha_{S,1}, \alpha_{S,2},$ $\alpha_{S,3},
	    \delta_{S,1},\ldots,\delta_{S,4k-3} \condition S\in \mathscr{S}\}$
and we define the sets 
\begin{eqnarray*}
\mathit{Beta} & = & \{\beta_b\condition b\in B\}, \\
\mathit{Zeta} & = & \{\zeta_{S}\condition S\in \mathscr{S}\}, \text{ and } \\
Q_S & = & \{\zeta_{S},\alpha_{S,1},\alpha_{S,2},\alpha_{S,3}, \delta_{S,1}, \ldots, \delta_{S,4k-3}\} \text{ for each $S\in \mathscr{S}$}.
\end{eqnarray*}
Figure~\ref{fig:SF-core-verification-min} shows the network of friends,
where
a dashed rectangle around a group of players means that all these players are friends of each other:
\begin{itemize}
	\item All players in $\mathit{Beta}$ are friends of each other.
	\item For every $S\in \mathscr{S}$, $\zeta_S$ is friend with every $\beta_b$ with $b\in S$ and
	with $\alpha_{S,1}$, $\alpha_{S,2}$, and $\alpha_{S,3}$.
	\item For every $S\in \mathscr{S}$, 
	$\alpha_{S,1}$, $\alpha_{S,2}$, $\alpha_{S,3}$, and $\delta_{S,1}$ are friends of each other.
	\item For every $S\in \mathscr{S}$, all players in $\{\delta_{S,1},\ldots, \delta_{S,4k-3}\}$ are friends of each other.
\end{itemize}

\begin{figure}[t] %
	\centering
	\begin{tikzpicture}
		\node (beta1) at (0,-0.9) {$\beta_1$};
		\node (betaweiter) at (0,-1.35) {$\vdots$};
		\node (betab) at (0,-2) {$\beta_b$};
		\node (betaweiter2) at (0,-2.45) {$\vdots$};
		\node (beta3k) at (0,-3.1) {$\beta_{3k}$};
		\node (zetaS1) at (2,-0.95) {$\zeta_{S_1}$} 
		edge ($(zetaS1)-(0:1cm)$)
		edge ($(zetaS1)-(15:1cm)$)
		edge ($(zetaS1)-(-15:1cm)$);
		\node (zetaweiter1) at (2,-1.3) {$\vdots$};
		\node (zetaSj) at (2,-2) {$\zeta_{S_j}$} 
		edge node [pos=0.4,above] {\scriptsize $b\in S_j$} (betab)
		edge (1,-2.25) edge (1,-2.5)
		edge ($(zetaSj)+(0:0.8cm)$)
		edge ($(zetaSj)+(12:0.8cm)$)
		edge ($(zetaSj)+(-12:0.8cm)$);
		\node (zetaweiter2) at (2,-2.5) {$\vdots$};
		\node (zetaS3k) at (2,-3.05) {$\zeta_{S_{3k}}$}
		edge ($(zetaS3k)-(0:1cm)$)
		edge ($(zetaS3k)-(15:1cm)$)
		edge ($(zetaS3k)-(-15:1cm)$);
		\node (alphaS11) at (3.5,-0.5) {$\alpha_{S_1,1}$} edge (zetaS1);
		\node (alphaS12) at (3.5,-0.95) {$\alpha_{S_1,2}$} edge (zetaS1);
		\node (alphaS13) at (3.5,-1.4) {$\alpha_{S_1,3}$} edge (zetaS1);
		\node (deltaS11) at (5,-0.95) {$\delta_{S_1,1}$} edge (alphaS11) edge (alphaS12) edge (alphaS13);
		\node (deltaweiter1) at (5.9,-0.95) {$\dots$};
		\node (deltaS14k) at (7,-0.95) {$\delta_{S_1,4k-3}$};
		\node (alphaS3k1) at (3.5,-2.6) {$\alpha_{S_{3k},1}$} edge (zetaS3k);
		\node (alphaS3k2) at (3.5,-3.05) {$\alpha_{S_{3k},2}$} edge (zetaS3k);
		\node (alphaS3k3) at (3.5,-3.5) {$\alpha_{S_{3k},3}$} edge (zetaS3k);
		\node (deltaS3k1) at (5,-3.05) {$\delta_{S_{3k},1}$} edge (alphaS3k1) edge (alphaS3k2) edge (alphaS3k3);
		\node (deltaweiter3k) at (5.9,-3.05) {$\dots$};
		\node (deltaS3k4k) at (7,-3.05) {$\delta_{S_{3k},4k-3}$};
		\node (weiter) at (3.5,-1.9) {$\vdots$};
		\node (dweiter) at (5.9,-1.9) {$\vdots$};
		\draw [dashed, rounded corners=3pt] ($(beta1)+(-0.5,0.25)$) rectangle ($(beta3k)+(0.5, -0.25)$);
		\draw [dashed, rounded corners=3pt] ($(alphaS11)+(-0.65,0.2)$) rectangle ($(alphaS13)+(0.65, -0.15)$);
		\draw [dashed, rounded corners=3pt] ($(alphaS3k1)+(-0.65,0.2)$) rectangle ($(alphaS3k3)+(0.65, -0.15)$);
		\node (E) [draw, dashed, rounded corners=3pt, fit={(deltaS11) (deltaS14k)}] {};
		\node (E) [draw, dashed, rounded corners=3pt, fit={(deltaS3k1) (deltaS3k4k)}] {};
		\node (Beta) at (0,-0.45) {$\mathit{Beta}$};
		\node (Zeta) at (2,-0.45) {$\mathit{Zeta}$};
		
		\node (QS1) at (8.5,-0.95) {$Q_{S_1}$};
		\node (Qweiter) at (8.5,-1.9) {$\vdots$};
		\node (QS3k) at (8.5,-3.05) {$Q_{S_{3k}}$};
	\end{tikzpicture}
	\caption{\label{fig:SF-core-verification-min}Network of friends
		in the proof of Theorem~\ref{thm:core-ver-in-conp} 
		that is used to show \conp-hardness of 
		core stability verification in min-based SF ACFGs.
		A dashed rectangle around a group of players indicates that all these 
		players are friends of each other.}
\end{figure}
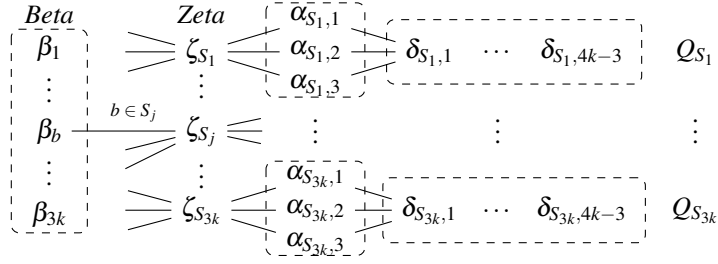
Furthermore, consider the coalition structure $\Gamma=
\{\mathit{Beta}, Q_{S_1},\ldots,Q_{S_{3k}}\}
$.
We will now show that
$\mathscr{S}$ contains an exact cover for $B$ if and only if $\Gamma$ is not core stable under the min-based SF model.

\proofonlyif Assume that 
there is an exact cover $\mathscr{S}'\subseteq\mathscr{S}$ for $B$. 
Then $\vert\mathscr{S}'\vert=k$. 
Consider coalition $C=\mathit{Beta}\cup\{\zeta_S\condition S\in \mathscr{S}'\}$.
$C$ blocks $\Gamma$, i.e., $\Gamma_{C\rightarrow\emptyset} \succsfmin_i \Gamma$ for all $i\in C$,
because 
	(a)~every $\beta_b\in \mathit{Beta}$ has $3k$ friends in $C$ but only $3k-1$ friends in $\mathit{Beta}$
	and 
	(b)~every $\zeta_S$ with $S\in \mathscr{S}'$ has $3$ friends and $4k-4$ enemies in $C$ but $3$ friends and $4k-3$ enemies in~$%
	Q_S$.

\proofif Assume that $\Gamma$ is not core stable
and let $C\subseteq N$ be a coalition that blocks~$\Gamma$.
Then $\Gamma_{C\rightarrow\emptyset} \succsfmin_i \Gamma$ for all $i\in C$.
First, observe that every $i\in C$ needs to have at least as many friends in $C$ as in $\Gamma(i)$. 
So, if any 
$\alpha_{S,j}$ %
or 
$\delta_{S,j}$ %
is in $C$, 
it follows quite directly that $Q_S\subseteq C$.
However, since $Q_S$ is a coalition in $\Gamma$ and 
since every other player (from $N\setminus Q_S$) is an enemy of all $\delta$-players, 
any coalition $C$ with $Q_S\subseteq C$ 
cannot be a blocking coalition for~$\Gamma$.
This contradiction implies that
no 	$\alpha_{S,j}$ or $\delta_{S,j}$ is in~$C$.

We now have $C\subseteq \mathit{Beta}\cup \mathit{Zeta}$.
Since any $\beta_b\in C$ has $3k-1$ friends and no enemies in~$\Gamma(\beta_b)$ and prefers $\Gamma_{C\rightarrow\emptyset}$ to~$\Gamma$, one of the following holds: %
	(a) $\beta_b$ has at least $3k$ friends in $C$ or
	(b) $\beta_b$ has $3k-1$ friends and no enemies in $C$
	and $\beta_b$'s friends assign a higher value to $\Gamma_{C\rightarrow\emptyset}$ than to~$\Gamma$. 
For a contra\-diction, assume that (b) holds for some $\beta_b\in C$.
First, observe that there are exactly $3k$ players in $C$ (namely,
$\beta_b$ and $\beta_b$'s $3k-1$ friends).
We now distinguish two cases:

\smallskip\textbf{Case~1:} \emph{All the $3k-1$ friends of $\beta_b$'s are $\beta$-players.} 
Then $C$ consists of all $\beta$-players, i.e., $C=\mathit{Beta}$. This is a contradiction, as $\mathit{Beta}$ is already a coalition in~$\Gamma$.

\smallskip\textbf{Case~2:} \emph{There are some $\zeta$-players in $C$ that are $\beta_b$'s friends.} 
Since $\beta_b$ has three $\zeta$-friends in total and no enemies in~$C$, there are between one and three $\zeta$-players in~$C$.
Hence, there are between $3k-3$ and $3k-1$ $\beta$-players in~$C$. 
Then one of the $\beta$-players has no $\zeta$-friend in~$C$. 
(The at most three $\zeta$-players are friends with at most nine $\beta$-players, but $3k-3>9$ for $k>4$.)
Consequently, this $\beta$-player has 
only the other (at most $3k-2$) $\beta$-players as friends in~$C$
and does not prefer $\Gamma_{C\rightarrow\emptyset}$ to $\Gamma$. This is a~contradiction.
\smallskip

Hence, option (a) holds for each $\beta_b\in C$. 
In total, each $\beta_b$ has exactly three \mbox{$\zeta$-friends} and $3k-1$ $\beta$-friends. 
Thus at least $3k-3$ of $\beta_b$'s friends in $C$ are $\beta$-players
and at least one of $\beta_b$'s friends in $C$ is a $\zeta$-player.
Also counting $\beta_b$ herself, there are at least $3k-2$ $\beta$-players in~$C$.
Since all of these $3k-2$ $\beta$-players 
have at least one $\zeta$-friend in~$C$, 
there are at least $k$ $\zeta$-players in~$C$.
(Note that $k-1$ $\zeta$-players are friends with at most
$3k-3$ $\beta$-players.)

Consider some $\zeta_S\in C$. 
Since $\zeta_S$ has three friends and $4k-3$ enemies in~$Q_S$,
at most three friends in~$C$,
and prefers $\Gamma_{C\rightarrow\emptyset}$ to~$\Gamma$, 
$\zeta_S$ has exactly three friends and at most $4k-3$ enemies in~$C$. 
Hence, $C$ contains at most $4k-3+3+1=4k+1$ players.

So far we know that there are at least $3k-2$ $\beta$-players
in~$C$.
If $C$ contains exactly $3k-2$ (or $3k-1$) $\beta$-players then each of this players has only $3k-3$ (or $3k-2$) $\beta$-friends in $C$ and additionally needs at least three (or two) $\zeta$-friends in~$C$.
Hence, we have at least $(3k-2)\cdot 3= 9k-6$ (or $%
6k-2$) edges between the $\beta$- and $\zeta$-players in~$C$.
Then there are at least $3k-2$ (or $2k$) $\zeta$-players in~$C$.
Thus there are at least $(3k-2)+(3k-2)=6k-4$ (or $%
5k-1$) players in $C$ which is a contradiction %
because there are at most $4k+1$ players in~$C$.
Hence, there are exactly $3k$ $\beta$-players in~$C$.

Summing up, there are  
exactly $3k$ $\beta$-players,
at least $k$ $\zeta$-players,
and at most $4k+1$ players in~$C$.
Hence, there are $k$ or $k+1$ $\zeta$-players in~$C$.
For the sake of contradiction,
assume that there are $k+1$ $\zeta$-players in~$C$.
Then each $\zeta_S \in C$ has $4k-3$ enemies in~$C$. 
Since $\zeta_S$ prefers $\Gamma_{C\rightarrow\emptyset}$ to~$\Gamma$,
this implies that $\zeta_S$ has exactly three friends and $4k-3$ enemies in $C$ and 
the minimal value assigned to $\Gamma_{C\rightarrow\emptyset}$ by $\zeta_S$'s friends is higher than the minimal value assigned to~$\Gamma$ by $\zeta_S$'s friends.
In both coalition structures, the minimal value is given by $\zeta_S$'s $\alpha$-friends. %
However, since these $\alpha$-players lose $\zeta_S$ as a friend when $\zeta_S$ deviates to~$C$, the minimal value assigned to $\Gamma$ is higher than for $\Gamma_{C\rightarrow\emptyset}$. This is a contradiction.
Hence, there are exactly $k$ $\zeta$-players in~$C$.
Finally, since every of the $3k$ $\beta_b\in C$ has one of the $k$ $\zeta_{S}\in C$ as a friend, it holds that $\{S \condition \zeta_{S}\in C \}$ is an exact cover for~$B$.
This completes the \conp-hardness proof for min-based SF ACFGs.
\medskip

For sum-based SF ACFGs, \conp-hardness of core stability verification can
be shown by a similar construction.
Again, given an instance
$(B,\mathscr{S})$ of \rxthreec,
with $B=\{1,\dots,3k\}$, 
$\mathscr{S}=\{S_1,\dots, S_{3k}\}$, and $k>8$,
we construct the following ACFG. The set of players is 
$N= \{\beta_b\condition b\in B\}
\cup \{\zeta_S, \alpha_{S,1}, \alpha_{S,2}, \alpha_{S,3},
\delta_{S,1},\ldots, \delta_{S,4k-3} \condition S\in \mathscr{S}\}$.
We define the sets $Beta=\{\beta_b\condition b\in B\}$ %
and $Q_S= \{%
\alpha_{S,1},\alpha_{S,2},\alpha_{S,3},\delta_{S,1},\ldots,\delta_{S,4k-3}\}$ for each $S\in \mathscr{S}$.
The network of friends is given in Figure \ref{fig:SF-core-verification},
where
a dashed rectangle around a group of players means that all these players are friends of each other:
\begin{itemize}
	\item All players in $Beta$ are friends of each other.
	\item For every $S\in \mathscr{S}$, all players in
	$Q_S$
	are friends of each other.
	\item For every $S\in \mathscr{S}$, $\zeta_S$ is friend with
	$\alpha_{S,1}$, $\alpha_{S,2}$, and $\alpha_{S,3}$
	and with every $\beta_b$ with $b\in S$.
\end{itemize}

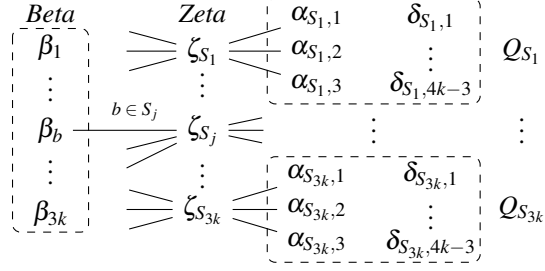
\begin{figure}
	\centering
	\begin{tikzpicture}
		\node (beta1) at (0,-0.9) {$\beta_1$};
		\node (betaweiter) at (0,-1.35) {$\vdots$};
		\node (betab) at (0,-2) {$\beta_b$};
		\node (betaweiter2) at (0,-2.45) {$\vdots$};
		\node (beta3k) at (0,-3.1) {$\beta_{3k}$};
		\node (zetaS1) at (2,-0.95) {$\zeta_{S_1}$} 
		edge ($(zetaS1)-(0:1cm)$)
		edge ($(zetaS1)-(15:1cm)$)
		edge ($(zetaS1)-(-15:1cm)$);
		\node (zetaweiter1) at (2,-1.3) {$\vdots$};
		\node (zetaSj) at (2,-2) {$\zeta_{S_j}$} 
		edge node [pos=0.4,above] {\scriptsize $b\in S_j$} (betab)
		edge (1,-2.25) edge (1,-2.5)
		edge ($(zetaSj)+(0:0.8cm)$)
		edge ($(zetaSj)+(12:0.8cm)$)
		edge ($(zetaSj)+(-12:0.8cm)$);
		\node (zetaweiter2) at (2,-2.5) {$\vdots$};
		\node (zetaS3k) at (2,-3.05) {$\zeta_{S_{3k}}$}
		edge ($(zetaS3k)-(0:1cm)$)
		edge ($(zetaS3k)-(15:1cm)$)
		edge ($(zetaS3k)-(-15:1cm)$);
		\node (alphaS11) at (3.5,-0.5) {$\alpha_{S_1,1}$} edge (zetaS1);
		\node (alphaS12) at (3.5,-0.95) {$\alpha_{S_1,2}$} edge (zetaS1);
		\node (alphaS13) at (3.5,-1.4) {$\alpha_{S_1,3}$} edge (zetaS1);
		\node (deltaS11) at (5,-0.5) {$\delta_{S_1,1}$}; 
		\node (deltaweiter1) at (5,-0.95) {$\vdots$};
		\node (deltaS14k) at (5,-1.4) {$\delta_{S_1,4k-3}$};
		\node (alphaS3k1) at (3.5,-2.6) {$\alpha_{S_{3k},1}$} edge (zetaS3k);
		\node (alphaS3k2) at (3.5,-3.05) {$\alpha_{S_{3k},2}$} edge (zetaS3k);
		\node (alphaS3k3) at (3.5,-3.5) {$\alpha_{S_{3k},3}$} edge (zetaS3k);
		\node (deltaS3k1) at (5,-2.6) {$\delta_{S_{3k},1}$};
		\node (deltaweiter3k) at (5,-3.05) {$\vdots$};
		\node (deltaS3k4k) at (5,-3.5) {$\delta_{S_{3k},4k-3}$};
		\node (weiter) at (4.25,-1.9) {$\vdots$};
		\draw [dashed, rounded corners=3pt] ($(beta1)+(-0.5,0.25)$) rectangle ($(beta3k)+(0.5, -0.25)$);
		\draw [dashed, rounded corners=3pt] ($(alphaS11)+(-0.65,0.25)$) rectangle ($(deltaS14k)+(0.65, -0.25)$);
		\draw [dashed, rounded corners=3pt] ($(alphaS3k1)+(-0.65,0.25)$) rectangle ($(deltaS3k4k)+(0.65, -0.25)$);
		\node (Beta) at (0,-0.45) {$\mathit{Beta}$};
		\node (Zeta) at (2,-0.45) {$\mathit{Zeta}$};
		
		\node (QS1) at (6.2,-0.95) {$Q_{S_1}$};
		\node (Qweiter) at (6.2,-1.9) {$\vdots$};
		\node (QS3k) at (6.2,-3.05) {$Q_{S_{3k}}$};
	\end{tikzpicture}
	\caption{\label{fig:SF-core-verification}Network of friends
		in the proof of Theorem~\ref{thm:core-ver-in-conp}
		that is used to show \conp-hardness of 
		core stability verification in sum-based SF ACFGs.
		A dashed rectangle around a group of players indicates that all these 
		players are friends of each other.}
\end{figure}

Similar arguments as above show that
the coalition structure $\Gamma=\{Beta\}\cup\{\{\zeta_{S}\}\cup Q_S\condition S\in \mathscr{S} \}$
is not core stable under sum-based SF preferences if and only if
$\mathscr{S}$ contains an exact cover for $B$.
\end{proofs}

\subsection{Popularity and Strict Popularity}
\label{subsec:POP}

Now we take a look at popularity and strict popularity. 
For all considered models of altruism,
there are games for which no (strictly) popular coalition structure exists.
\begin{example}
	\label{ex:no-strictly-popular}
	Let $N=\{1,\ldots,10\}$ and consider the network
	of friends shown in Figure~\ref{fig:no-strictly-popular}.
	\begin{figure}[t] %
		\centering
		\begin{tikzpicture}
		\node (1) at (0,0) {$1$};
		\node (2) at (1,0) {$2$};
		\node (3) at (2,0) {$3$};
		\node (4) at (3,0) {$4$};
		\node (5) at (4,0) {$5$};
		\node (6) at (5,0) {$6$};
		\node (7) at (6,0) {$7$};
		\node (8) at (4,-0.6) {$8$};
		\node (9) at (5,-0.6) {$9$};
		\node (10) at (6,-0.6) {$10$};
		
		\draw (1) -- (2);
		\draw (2) -- (3);
		\draw (3) -- (4);
		\draw (4) -- (5);
		\draw (5) -- (6);
		\draw (6) -- (7);
		\draw (4) -- (8);
		\draw (8) -- (9);
		\draw (9) -- (10);
		\end{tikzpicture}
		\caption{\label{fig:no-strictly-popular}
			Network of friends for Example~\ref{ex:no-strictly-popular} }
	\end{figure}
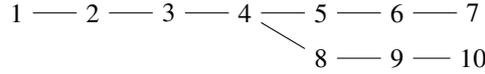
	Then there is no strictly popular and no popular coalition structure for any of the sum-based or min-based degrees of altruism.
	Since perfectness implies popularity, there is also no perfect coalition structure for this {ACFG}.

	Recall from Footnote~\ref{foo:bell} that there are $115,975$
        possible coalition structures for this game with ten players,
	which we all tested for this example by brute force.

\end{example}

We now show that, under sum-based and min-based SF preferences, it is
hard to verify if a given coalition structure is popular or strictly
popular, and it is also hard to decide whether there exists a strictly
popular coalition structure for a given SF ACFG.

\begin{theorem}\label{thm:SF-str-pop-verification}
  Popularity verification and strict popularity verification are
  in \conp\ for any ACFG. 
	For (sum-based and min-based) SF ACFGs, 
	popularity verification and strict popularity verification
        are \conp-complete
	and 
	strict popularity existence is \conp-hard.
\end{theorem}

\begin{proofs}
	First, we observe that the verification problems %
	are in \conp:
        To verify that a given coalition structure $\Gamma$ is not
        (strictly) popular, we can nondeterministically guess a
        coalition structure~$\Delta$, compare both coalition
        structures in polynomial~time, and accept exactly if $\Delta$
        is more popular than (or at least as popular as)~$\Gamma$.
		
	To show \conp-hardness of strict popularity verification for min-based SF ACFGs, 
	we again employ a polynomial-time many-one reduction from \rxthreec.
	Let~$(B,\mathcal{S})$
	be an instance of \rxthreec, consisting of a set $B=\{1,\ldots,3k\}$ and a collection $\mathcal{S}=\{S_1,\cdots,S_{3k}\}$ of 3-element subsets of $B$. Recall that every element of~$B$ occurs in exactly three sets in~$\mathcal{S}$ and the question is whether there is an exact cover $\mathcal{S}'\subseteq\mathcal{S}$ of~$B$.
	
	We now construct a network of friends based on this instance. The set of players is given by 
	$N=\{\alpha_1,\ldots,\alpha_{2k}\}\cup\{\beta_b \condition b\in B\}\cup\{\zeta_S,\eta_{S,1},\eta_{S,2} \condition S \in\mathcal{S}\}$, so in total we have $n=14k$ players. For convenience, we define %
	$\mathit{Alpha}=\{\alpha_1,\ldots,\alpha_{2k}\}$, $\mathit{Beta}=\{\beta_b \condition b\in B\}$, and $Q_S =\{\zeta_S,\eta_{S,1},\eta_{S,2} \condition S \in\mathcal{S}\}$ for $S\in\mathcal{S}$. The network of friends is shown in Figure~\ref{fig:StrictPopVeriproof}, where a dashed square around a group of players means that all these players are friends of each other:
	All players in $\mathit{Alpha} \cup  \mathit{Beta}$ are friends of each other;
	for every $S\in\mathcal{S}$, all players in $Q_S$ are friends of each other; and
	$\zeta_S$ is a friend of every $\beta_b$ with $b\in S$.
	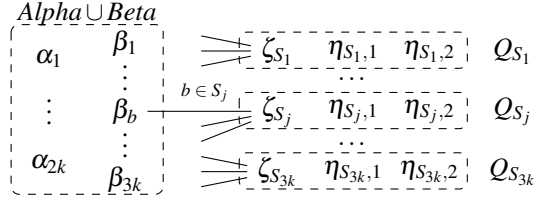
\begin{figure}[t]
		\centering
		\begin{tikzpicture} %
			\node (alpha1) at (1,-1.1) {$\alpha_1$};
			\node (alphaweiter) at (1,-1.7) {$\vdots$};
			\node (alpha2k) at (1,-2.5) {$\alpha_{2k}$};
			\node (beta1) at (2,-0.9) {$\beta_1$};
			\node (betaweiter) at (2,-1.25) {$\vdots$};
			\node (betab) at (2,-1.8) {$\beta_b$};
			\node (betaweiter2) at (2,-2.15) {$\vdots$};
			\node (beta3k) at (2,-2.7) {$\beta_{3k}$};
			\node (zetaS1) at (4,-1) {$\zeta_{S_1}$} 
			edge (3,-0.8) edge (3,-1) edge (3,-1.2);
			\node (zetaSj) at (4,-1.8) {$\zeta_{S_j}$} 
			edge node [pos=0.45,above] {\scriptsize $b\in S_j$} (betab)
			edge (3,-2) edge (3,-2.2);
			\node (zetaS3k) at (4,-2.6) {$\zeta_{S_{3k}}$}
			edge (3,-2.8) edge (3,-2.6) edge (3,-2.4);
			\node (etaS11) at (5,-1) {$\eta_{S_1,1}$}; %
			\node (etaS12) at (6,-1) {$\eta_{S_1,2}$}; %
			\node (etaweiter1) at (5,-1.37) {$\dots$};
			\node (etaSj1) at (5,-1.8) {$\eta_{S_j,1}$}; %
			\node (etaSj2) at (6,-1.8) {$\eta_{S_j,2}$}; %
			\node (etaweiter2) at (5,-2.23) {$\dots$}; %
			\node (etaS3k1) at (5,-2.6) {$\eta_{S_{3k},1}$}; %
			\node (etaS3k2) at (6,-2.6) {$\eta_{S_{3k},2}$}; %
			\draw [dashed, rounded corners=3pt] ($(zetaS1)+(-0.5,0.25)$) rectangle ($(etaS12)+(0.5, -0.23)$);
			\draw [dashed, rounded corners=3pt] ($(zetaSj)+(-0.5,0.25)$) rectangle ($(etaSj2)+(0.5, -0.23)$);
			\draw [dashed, rounded corners=3pt] ($(zetaS3k)+(-0.5,0.25)$) rectangle ($(etaS3k2)+(0.5, -0.23)$);
			\draw [dashed, rounded corners=3pt] ($(alpha1)+(-0.5,0.45)$) rectangle ($(beta3k)+(0.5, -0.2)$);
			\node (QS1) at (7.1,-1) {$Q_{S_1}$};
			\node (QSj) at (7.1,-1.8) {$Q_{S_j}$};
			\node (QS3k) at (7.1,-2.6) {$Q_{S_{3k}}$};
			\node (AlphaBeta) at (1.5,-0.5) {$\mathit{Alpha} \cup \mathit{Beta}$};
		\end{tikzpicture}
		\centering
		\caption{Network of friends in the proof of Theorem~\ref{thm:SF-str-pop-verification}
		that is used to show \conp-hardness of 
		strict popularity verification in min-based SF ACFGs.
		A dashed rectangle around a group of players indicates that all these 
		players are friends of each other.}
		\label{fig:StrictPopVeriproof}
	\end{figure} 
	
	We consider the coalition structure $\Gamma=\{\mathit{Alpha} \cup \mathit{Beta}\}\cup\{Q_S \condition S\in\mathcal{S}\}$ and will now show that 
	$\mathcal{S}$ contains an exact cover for $B$ if and only if $\Gamma$ is not strictly popular
	under min-based SF preferences.
	
	\proofonlyif Assuming that there is an exact cover $\mathcal{S}'\subset\mathcal{S}$ for~$B$, we define the coalition structure $\Delta = \{\mathit{Alpha}\,\cup\,\mathit{Beta}\, \cup\, \bigcup_{S\in\mathcal{S}'} Q_S\} \cup \{Q_S \condition S\in\mathcal{S}\setminus\mathcal{S}'\}$. We will now show that $\Delta$ is as popular as $\Gamma$ under min-based SF preferences.
	
	First, all $2k$ $\alpha$-players prefer $\Gamma$ to~$\Delta$, since they only add enemies to their coalition in~$\Delta$.
	Second,
	the $3k$ $\beta$-players prefer $\Delta$ to~$\Gamma$, as each $\beta$-player gains a $\zeta$-friend and then has $5k$ friends instead of $5k-1$.
	Next,
	we consider the $Q_S$-groups for $S\in\mathcal{S}'$, i.e., the groups that were added to $\mathit{Alpha}\cup \mathit{Beta}$ in $\Delta$. We observe that every $\zeta_S$-player in these $Q_S$-groups prefers $\Delta$ to~$\Gamma$, since $\zeta_S$ gains three additional $\beta$-friends. For the $\eta$-players, on the other hand, the new coalition only contains more enemies, so the $\eta$-players prefer $\Gamma$ to~$\Delta$. Since we have $\vert\mathcal{S}'\vert=k$, this means $k$ $\zeta$-players prefer $\Delta$ to~$\Gamma$, and $2k$ $\eta$-players prefer $\Gamma$ to~$\Delta$.
	Finally, 
	we consider the remaining $Q_S$-groups with $S\in\mathcal{S}\setminus\mathcal{S}'$. Here, the coalition containing these players is the same in $\Gamma$ and $\Delta$. 
	Hence,
	for each player $p\in Q_S$, we have %
	$v_p(\Gamma) = v_p(\Delta)$.
	Thus the players have to ask their friends for their valuations. For $\zeta_S\in Q_S$ with $S\in\mathcal{S}\setminus\mathcal{S}'$, the minimum value of her %
	friends is in both structures given by an $\eta$-friend, since $\eta_{S,1}$ and $\eta_{S,2}$ value $\Gamma$ and $\Delta$ both with $n\cdot2$, while the $\beta$-friends of $\zeta_S$ assign values $n\cdot(5k-1)$ to $\Gamma$ and $n\cdot5k-(3k-1)$ to~$\Delta$. So we have $\utilitysfmin_{\zeta_S}(\Gamma) = \utilitysfmin_{\zeta_S}(\Delta)$ and, therefore, $2k$ $\zeta$-players that are indifferent. The $\eta$-players in~$Q_S$, $S\in\mathcal{S}\setminus\mathcal{S}'$, are also indifferent, as all their friends value $\Gamma$ and $\Delta$ the same.
	In total,
        $\#_{\Delta\succ\Gamma} = \vert \mathit{Beta}\,\cup\,\{\zeta_S \condition S\in\mathcal{S}'\}\vert = 4k = \vert{\mathit{Alpha}}\,\cup\,\{\eta_{S,1}, \eta_{S,2} \condition S\in\mathcal{S}'\}\vert = \#_{\Gamma\succ\Delta}$ and, therefore, $\Delta$ is exactly as popular as~$\Gamma$, so $\Gamma$ is not strictly popular.
	
	\proofif Assuming that $\Gamma$ is not strictly popular, there is some coalition structure 
	$\Delta \in\mathcal{C}_N$ with $\Delta \neq \Gamma$ such that $\Delta$ is at least as popular as $\Gamma$ under min-based SF preferences. 
	We will now show that 
	this
	implies the existence of an exact cover for $B$ in~$\mathcal{S}$.
	
	First of all, we observe that all $\alpha$-players' most preferred coalition is $\mathit{Alpha}\cup \mathit{Beta}$, as it contains all their friends and no enemies. 
	Thus we have 
	$\Gamma \succsfmin_\alpha \Delta$ if ${\mathit{Alpha}\cup \mathit{Beta}} \notin\Delta$ and $\Gamma \simsfmin_\alpha \Delta$ if ${\mathit{Alpha} \cup \mathit{Beta}} \in\Delta$.
	
	For the sake of contradiction,
	we assume that ${\mathit{Alpha}\cup \mathit{Beta}}\in\Delta$. As $\Delta\neq\Gamma$, the players in the $Q_S$-groups have to be partitioned differently. However, 
	that would not increase any player's valuation
	since every player in $Q_S$ can only lose friends and gain enemies. That means that no $\beta$-player prefers $\Delta$ to~$\Gamma$, as they are in the same coalition as in $\Gamma$ and their friends are not more satisfied.
	We also have at least three players of a $Q_S$-group that are no longer in the same coalition, so they prefer $\Gamma$ to~$\Delta$. This is a contradiction, 
	as we assumed that $\Delta$ is at least as popular as $\Gamma$. 
	Thus we have ${\mathit{Alpha}\cup \mathit{Beta}} \notin \Delta$.
	
	Now consider the $\eta$-players. For every $S\in\mathcal{S}$, we know that $Q_S$ is the best valued coalition for $\eta_{S,1}$ and~$\eta_{S,2}$. So again, $\eta_{S,1}$ and $\eta_{S,2}$ prefer $\Gamma$ to $\Delta$ if and only if $Q_S\notin\Delta$, and they are indifferent otherwise.  Define $k' = \vert\{S \in\mathcal{S} \condition Q_S \notin \Delta\}\vert$. So $2k'$ is the number of $\eta$-players that prefer $\Gamma$ to~$\Delta$, and
	the remaining $6k-2k'$ $\eta$-players are indifferent between $\Gamma$ and~$\Delta$. 
	We first collect some observations:
	\begin{enumerate}
		\item All $2k$ $\alpha$-players prefer $\Gamma$ to~$\Delta$.
		\item $2k'$ $\eta$-players prefer $\Gamma$ to~$\Delta$, and $6k-2k'$ $\eta$-players are indifferent.
		\item $3k-k'$ $\zeta$-players are in the same coalition in both coalition structures, so 
		their utilities depend on their friends' valuations.
		In~$\Gamma$, the minimum value of their friends is given by an $\eta$-player. %
		Since this $\eta$-player is also in the same coalition in $\Delta$ and thus assigns the same value, it is not possible that the minimum value of the friends 
		is higher in $\Delta$ than in~$\Gamma$.
		So $3k-k'$ $\zeta$-players are indifferent or prefer $\Gamma$ to~$\Delta$.
		\item We have $14k$ players in total, so we can have at most $14k-2k-2k'-(6k-2k')-(3k-k')=3k+k'$ players that prefer $\Delta$ to~$\Gamma$.
	\end{enumerate}
	
	Next, we show that $k'=k$. First, assume that $k'>k$: We have $\#_{\Gamma\succ\Delta} \geq 2k+2k'$, and since $k'>k$, we have $2k+2k'>3k+k'\geq \#_{\Delta\succ\Gamma}$. This is a contradiction to $\#_{\Gamma\succ\Delta}\leq\#_{\Delta\succ\Gamma}$, so we obtain $k'\leq k$.
	
	Second, let us assume $k'<k$: Since every $\zeta$-player has three $\beta$-friends and there are $k'$ $\zeta$-players that are not in their respective $Q_S$ coalition in $\Delta$, there are at most $3k'$ $\beta$-players that gain a $\zeta$-friend in $\Delta$. The $3k-3k'$ other $\beta$-players have at most $5k-1$ friends in~$\Delta$, namely all other $\alpha$- and $\beta$-players. But as ${\mathit{Alpha}\cup \mathit{Beta}}\notin\Delta$, they would also gain at least one enemy, so we have $3k-3k'$ $\beta$-players that prefer $\Gamma$. That means we have $\#_{\Gamma\succ\Delta}\geq 2k+2k'+3k-3k' = 5k-k'$ and $\#_{\Delta\succ\Gamma} \leq 3k+k' - (3k-3k') = 4k'$. Since $k'<k$, we have $5k-k'>5k-k = 4k > 4k'$, and therefore, $\#_{\Gamma\succ\Delta} > \#_{\Delta\succ\Gamma}$, which again is a contradiction. Thus we conclude that $k'\geq k$ and, in total, $k'=k$.
	
	Consequently, we know that $4k$ players prefer $\Gamma$ to~$\Delta$, namely all $\alpha$-players and the $2k$ $\eta$-players that are not in $Q_S$ anymore. Subtracting all the indifferent players, we observe that all other players have to prefer $\Delta$ to $\Gamma$ in order to ensure $\#_{\Gamma\succ\Delta}\leq\#_{\Delta\succ\Gamma}$. These other players are the $3k$ $\beta$-players and the $k$ $\zeta$-players that are not in $Q_S$ anymore. Finally, that is only possible if every $\beta$-player gains a $\zeta$-friend in~$\Delta$. Hence each one of those $k$ $\zeta$-players has to be friends with three different $\beta$-players. Therefore, the set $\{S \in\mathcal{S}\condition Q_S \notin \Delta\}$ is an exact cover for $B$.
\medskip	

	To show \conp-hardness of strict popularity verification for sum-based SF ACFGs, 
	we use a similar construction.
	For
	an instance $(B,\mathscr{S})$ of \rxthreec\ with $B=\{1,\dots,3k\}$ and 
	$\mathscr{S}=\{S_1,\dots, S_{3k}\}$, where each element of $B$ occurs 
	in exactly three sets in~$\mathscr{S}$,
	we construct the following ACFG. The set of players is given by
	$N=\{\alpha_1,\ldots,\alpha_{5k}\}\cup \{\beta_b \condition b\in B\}
	\cup \{\zeta_S, \eta_S \condition S\in \mathscr{S}\}$.
	Let $Alpha=\{\alpha_1,\ldots,\alpha_{5k}\}$, $Beta=\{\beta_b\condition b\in B\}$, and $Q_S= \{\zeta_{S},\eta_{S}\}$ for each $S\in \mathscr{S}$.
	The network of friends is given in Figure~\ref{fig:SF-pop-verification},
	where
	a dashed rectangle around a group of players means that all these players are friends of each other:
	All players in %
		$Alpha\cup Beta$
		are friends of each other and,
	for every $S\in \mathscr{S}$, 
	$\zeta_S$ is friends with $\eta_S$ and every $\beta_b$ with $b\in S$.
	
	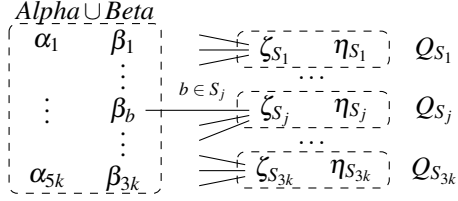
\begin{figure}[t] %
		\centering
		\begin{tikzpicture} %
			\node (alpha1) at (1,-0.9) {$\alpha_1$};
			\node (alphaweiter) at (1,-1.7) {$\vdots$};
			\node (alpha2k) at (1,-2.7) {$\alpha_{5k}$};
			\node (beta1) at (2,-0.9) {$\beta_1$};
			\node (betaweiter) at (2,-1.25) {$\vdots$};
			\node (betab) at (2,-1.8) {$\beta_b$};
			\node (betaweiter2) at (2,-2.15) {$\vdots$};
			\node (beta3k) at (2,-2.7) {$\beta_{3k}$};
			\node (zetaS1) at (4,-1) {$\zeta_{S_1}$} 
			edge (3,-0.8) edge (3,-1) edge (3,-1.2);
			\node (zetaSj) at (4,-1.8) {$\zeta_{S_j}$} 
			edge node [pos=0.45,above] {\scriptsize $b\in S_j$} (betab)
			edge (3,-2) edge (3,-2.2);
			\node (zetaS3k) at (4,-2.6) {$\zeta_{S_{3k}}$}
			edge (3,-2.8) edge (3,-2.6) edge (3,-2.4);
			\node (etaS11) at (5,-1) {$\eta_{S_1}$}; %
			\node (etaweiter1) at (4.5,-1.37) {$\dots$};
			\node (etaSj1) at (5,-1.8) {$\eta_{S_j}$}; %
			\node (etaweiter2) at (4.5,-2.23) {$\dots$}; %
			\node (etaS3k1) at (5,-2.6) {$\eta_{S_{3k}}$}; %
			\draw [dashed, rounded corners=3pt] ($(zetaS1)+(-0.5,0.25)$) rectangle ($(etaS11)+(0.5, -0.23)$);
			\draw [dashed, rounded corners=3pt] ($(zetaSj)+(-0.5,0.25)$) rectangle ($(etaSj1)+(0.5, -0.23)$);
			\draw [dashed, rounded corners=3pt] ($(zetaS3k)+(-0.5,0.25)$) rectangle ($(etaS3k1)+(0.5, -0.23)$);
			\draw [dashed, rounded corners=3pt] ($(alpha1)+(-0.5,0.25)$) rectangle ($(beta3k)+(0.5, -0.2)$);
			\node (QS1) at (6.1,-1) {$Q_{S_1}$};
			\node (QSj) at (6.1,-1.8) {$Q_{S_j}$};
			\node (QS3k) at (6.1,-2.6) {$Q_{S_{3k}}$};
			\node (AlphaBeta) at (1.5,-0.5) {$\mathit{Alpha} \cup \mathit{Beta}$};
		\end{tikzpicture}
		\caption{\label{fig:SF-pop-verification}Network of friends
			in the proof of Theorem~\ref{thm:SF-str-pop-verification}
			that is used to show \conp-hardness of 
			strict popularity verification in sum-based SF ACFGs.
			A dashed rectangle around a group of players indicates that all these 
			players are friends of each other.}
	\end{figure}
	
	Consider the coalition structure 
	$\Gamma=\{Alpha\cup Beta,Q_{S_1},\ldots,Q_{S_{3k}} \}$.
	We show that $\mathscr{S}$ contains an exact cover for $B$ if and only if $\Gamma$ is not strictly popular.
	
	\proofonlyif Assuming that there is an exact cover $\mathscr{S}'\subseteq\mathscr{S}$ for $B$
	and considering coalition structure
	$\Delta=\{Alpha\cup Beta\cup \bigcup_{S\in \mathscr{S}'} Q_S \} \cup \{ Q_S \condition S\in \mathscr{S}\setminus \mathscr{S}'\}$,
	it can be shown with similar arguments as before that
	$\#_{\Delta \succ \Gamma} =
	\vert \{\beta_1,\ldots,\beta_{3k},\zeta_{S_1},\ldots,\zeta_{S_{3k}} \}\vert 
	=6k=
	\vert \{\alpha_1,\ldots,\alpha_{5k}\}\cup\{\eta_{S}\condition S\in \mathscr{S}'\} \vert 
	=\#_{\Gamma \succ \Delta}$. 
	Hence, $\Delta$ and $\Gamma$ are equally popular.

	\proofif Assuming that $\Gamma$ is not strictly popular, i.e., that there is a coalition structure $\Delta\in \coalstr$, $\Delta\neq \Gamma$, with
	$\#_{\Gamma \succ \Delta} \leq \#_{\Delta \succ \Gamma}$,
	it can be shown similarly as before that the set
	$\{S\in \mathscr{S}\condition Q_S \notin \Delta\}$ 
	is an exact cover for~$B$.

\medskip
        The results for strict popularity existence
	and popularity verification can be shown by
	slightly modifying the above reductions.
\medskip
	
	To show that strict popularity existence is \conp-hard for min-based and sum-based SF ACFGs,
	we consider the same two reductions as before but the coalition structures $\Gamma$ are not given as a part of the problem instances. 
	Then, there is an exact cover for $B$ if and only if there is no strictly popular coalition structure. %
	In particular, if there is an exact cover for $B$,
	$\Gamma$ and $\Delta$ as defined in the proofs above are in a tie and every other coalition structure is beaten by $\Gamma$. 
	And if there is no exact cover for $B$ then
	$\Gamma$ beats every other coalition structure and thus is strictly popular.
\medskip

	Popularity verification for min-based and sum-based SF ACFGs can be shown
	to be \conp-complete by using the same constructions as for strict popularity verification 
	(see Figure~\ref{fig:StrictPopVeriproof} and~\ref{fig:SF-pop-verification}) but
	reducing the numbers of $\alpha$-players to $2k-1$ and $5k-1$, respectively.
	Then
	there is an exact cover for $B$ 
	if and only if $\Gamma$, as defined above, is not popular.~\end{proofs}

\subsection{Perfectness}
\label{subsec:PER}
Turning now to perfectness, we start with
the SF model.

\begin{theorem}\label{perfect_ver_SF}
	For any sum-based or min-based SF ACFG $(N,\succeq)$
	with an underlying network of friends~$G$,
	a coalition structure
	$\Gamma\in \coalstr$ is perfect 
	if and only if 
	it consists of the connected components of~$G$ and all of them
	are cliques.
\end{theorem}

\begin{proofs}
	From left to right, assume that the coalition structure
	$\Gamma\in \coalstr$ is perfect.  It then holds for all agents
	$i\in N$ and all coalition structures $\Delta\in \coalstr$,
	$\Delta\neq \Gamma$, that $i$ weakly prefers $\Gamma$ to~$\Delta$. %
	It follows that
	$v_i(\Gamma)\geq v_i(\Delta)$ for all $\Delta\in \coalstr$,
	$\Delta\neq \Gamma$, and $i\in N$.
	Hence, every agent $i\in N$ has the maximal valuation $v_i(\Gamma)= n\cdot \vert F_i\vert$ and is together with all of her friends and none of her enemies. 
	This implies that each coalition in $\Gamma$ is a connected component and a clique.

	The implication from right to left is obvious.~\end{proofs}

Since it is easy to check this characterization, perfect coalition
structures can be verified in polynomial time for sum-based and min-based SF ACFGs.  
It follows directly from Theorem~\ref{perfect_ver_SF} that the
corresponding existence problem is also in~\pol.

\begin{corollary}
	For any sum-based or min-based SF ACFG $(N,\succeq)$
	with an underlying network of friends~$G$,
	there
	exists a perfect coalition structure if
	and only if all connected components of~$G$ are cliques.
\end{corollary}	

We further get the following upper bounds.

\begin{proposition}
\label{prop:perfectness}
	For any ACFG,
	perfectness verification is
	in \conp.
\end{proposition}
\begin{proofs}
	Consider any ACFG $(N,\succeq)$.
	A coalition structure $\Gamma\in \coalstr$ is not perfect if
	and only if there is an agent $i\in N$ and a coalition structure
	$\Delta\in \coalstr$ such that $\Delta \succ_{i} \Gamma$.  Hence, we
	can nondeterministically guess an agent $i\in N$ and a coalition
	structure $\Delta\in \coalstr$ and verify in polynomial time
	whether
	$\Delta \succ_{i} \Gamma$.~\end{proofs}

Furthermore, we initiate the characterization of perfectness in ACFGs.
The \emph{diameter} of a connected graph component is the greatest
distance between any two of its vertices.
For sum-based EQ ACFGs, we get the following implication. 

\begin{proposition}
	\label{lemma:EQ-perfect}  
	For any sum-based EQ ACFG
	with an underlying network of friends~$G$,
	it holds that
	if 
	a coalition structure $\Gamma$ is perfect for it, then
	$\Gamma$ consists of the connected components of~$G$ and all these components 
	have a diameter of at most two.
\end{proposition}

	\begin{proofs}
		We first show that, in a perfect coalition structure, all agents have to be together with all their friends.
		For the sake of contradiction, assume that $\Gamma$ is perfect but there are 
		$i, j\in N$ with $j\in F_i$ and $j\notin \Gamma(i)$.
		We distinguish two cases.
		
		\smallskip\textbf{Case~1:} \emph{All $f\in F_i\cap\Gamma(i)$ have a friend in $\Gamma(j)$.}
		Consider the coalition structure $\Delta$ that results from the union of $\Gamma(i)$ and $\Gamma(j)$, i.e.,
		$\Delta=\Gamma\setminus\{\Gamma(i),\Gamma(j)\}\cup\{\Gamma(i)\cup\Gamma(j) \}$.
		It holds that $i$ and all friends of $i$'s %
		either gain an additional friend in $\Delta$ or their coalition stays the same:
			First,
			$i$ keeps all friends from $\Gamma(i)$ and gets $j$ as an additional friend. Hence, $i$ has at least one friend more in $\Delta$ than in $\Gamma$ and we have $v_i(\Delta)>v_i(\Gamma)$.
			Second,
			all friends $f\in F_i\cap\Gamma(i)$ have a friend in $\Gamma(j)$ and therefore also gain at least one additional friend from the union of the two coalitions. Hence, $v_f(\Delta)>v_f(\Gamma)$ for all $f\in F_i\cap\Gamma(i)$.
			Third,
			all friends $f\in F_i\cap\Gamma(j)$ have $i$ as friend. Hence, they also gain one friend from the union. Thus $v_f(\Delta)>v_f(\Gamma)$ for all $f\in F_i\cap\Gamma(j)$.
			Finally,
			all $f\in F_i$ who are not in $\Gamma(i)$ or $\Gamma(j)$ value $\Gamma$ and $\Delta$ the same because their coalition is the same in both coalition structures. 
			Hence, $v_f(\Delta)=v_f(\Gamma)$ for all $f\in F_i$
			with $f\notin\Gamma(j)$ and $f\notin\Gamma(i)$.
		Summing up, we have 
		$\utilityeqsum_{i}(\Delta) > \utilityeqsum_{i}(\Gamma)$,
                so $i$ prefers $\Delta$ to~$\Gamma$, 
		which is a contradiction to $\Gamma$ being perfect.
		
		\smallskip\textbf{Case~2:} \emph{There is an $f\in F_i\cap\Gamma(i)$ who has no friends in $\Gamma(j)$.}
		Consider the coalition structure $\Delta$ that results from $j$ moving to $\Gamma(i)$, i.e.,
		$\Delta= \Gamma_{j\rightarrow \Gamma(i)}$.
		Let $k\in F_i\cap\Gamma(i)$ be one of the agents who have no friends in $\Gamma(j)$. Then 
		$v_k(\Delta)=v_k(\Gamma)-1$;
		$v_i(\Delta)=v_i(\Gamma)+n$;
		for all $f \in F_k\cap\Gamma(i)$, $f\neq i$, we have $v_f(\Delta)\geq v_f(\Gamma)-1$; and
		for all $f \in F_k, f\notin\Gamma(i)$ (and $f\notin \Gamma(j)$), we have $v_f(\Delta)= v_f(\Gamma)$.
		Hence,
		\begin{align*}
		\utilityeqsum_{k}(\Delta)&= \sum_{a \in F_k \cup \{k\}} v_a(\Delta)
		= \sum_{a \in F_k\cap\Gamma(i), a\neq i} v_a(\Delta) + 
		\sum_{a \in F_k\setminus\Gamma(i)} v_a(\Delta) 
		+ v_k(\Delta) +v_i(\Delta)\\
		&\geq \sum_{a \in F_k\cap\Gamma(i), a\neq i} v_a(\Gamma)-1 + 
		\sum_{a \in F_k\setminus\Gamma(i)} v_a(\Gamma) 
		+ v_k(\Gamma)-1 +v_i(\Gamma)+n\\
		&= \sum_{a \in F_k \cup \{k\}} v_a(\Gamma) 
		-(\vert F_k\cap\Gamma(i)\vert-1) -1 +n\\
		&= \utilityeqsum_{k}(\Gamma) - \underbrace{\vert F_k\cap\Gamma(i)\vert}_{<n} +n 
		 > \utilityeqsum_{k}(\Gamma).
		\end{align*}
		Therefore, $k$ prefers $\Delta$ to $\Gamma$, which again is a contradiction to $\Gamma$ being perfect. 
		\smallskip
                
		Next,
		assume that $\Gamma$ is perfect but there is a coalition $C$ in $\Gamma$ that 
		has a diameter greater than two.
		Then there are agents $i,j\in C$ with a distance greater than two. 
		Thus $j$ is an enemy of $i$'s and an enemy of all of $i$'s friends.
		It follows that $i$ prefers coalition structure $\Gamma_{j\rightarrow \emptyset}$
		to $\Gamma$, 
		which is a contradiction to $\Gamma$ being perfect.
		
		Summing up,
                in a perfect coalition structure $\Gamma$ for a
                sum-based EQ ACFG
		every agent is together with all her friends and
		every coalition in $\Gamma$ has a diameter of at most two. 
		Together this implies that 
		$\Gamma$ consists of the connected components of $G$ and 
		all these components have a diameter of at most two.~\end{proofs}

From Propositions~\ref{prop:perfectness} and~\ref{lemma:EQ-perfect},
we get the following corollary.
\begin{corollary}
	For sum-based EQ ACFGs, perfectness existence is in \conp.
\end{corollary}
\OMIT{
\begin{proofs}
	Let $G$ be the underlying network of friends for an ACFG $(N,\succeq)$.
	Let, furthermore, $\Gamma\in \coalstr$ be the coalition
	structure that consists of the connected components of~$G$.  From
	Lemma~\ref{lemma:EQ-perfect-implies-cc-and-2clan} we know that if a
	coalition structure does not consist of the connected components of
	$G$, it is not perfect.  Hence, $\Gamma$ is the only coalition
	structure which possibly is perfect.  Therefore, there exists a
	coalition structure that is perfect under equal treatment if and only
	if $\Gamma$ is perfect.  Since perfectness verification is in \conp,
	it follows from this equivalence that existence is in \conp.
\end{proofs}
}

However, Proposition~\ref{lemma:EQ-perfect} is not an equivalence. The converse does not hold, as the following example shows.
\begin{example}\label{ex:2clans-not-perfect}
	Consider the sum-based EQ ACFG $(N,\succeqequal)$ with %
	the network of friends~$G$ in Figure~\ref{fig:2clans-not-perfect}.
	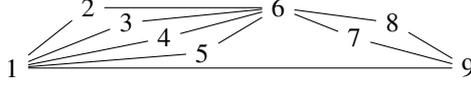
\begin{figure}[t] %
		\centering
		\begin{tikzpicture}
			\node (1) at (-3,0) {$1$};
			\node (2) at (-2,0.8) {$2$};
			\node (3) at (-1.5,0.6) {$3$};
			\node (4) at (-1,0.4) {$4$};
			\node (5) at (-0.5,0.2) {$5$};
			\node (6) at (3,0) {$9$};
			\node (7) at (1.5,0.4) {$7$};
			\node (8) at (2,0.6) {$8$};
			\node (9) at (0.5,0.8) {$6$};
			
			\draw (1) -- (2);
			\draw (1) -- (3);
			\draw (1) -- (4);
			\draw (1) -- (5);
			\draw (1) -- (6);
			\draw (2) -- (9);
			\draw (3) -- (9);
			\draw (4) -- (9);
			\draw (5) -- (9);
			\draw (6) -- (7);
			\draw (6) -- (8);
			\draw (7) -- (9);
			\draw (8) -- (9);
		\end{tikzpicture}
		\caption{\label{fig:2clans-not-perfect}
			Network of friends for
			Example~\ref{ex:2clans-not-perfect}}
	\end{figure}
	The coalition structure
	$\Gamma=\{N\}$ consists of the only connected component
	of $G$, which 
	has a diameter of two.
	However, 
	agent $1$ prefers $\Delta=\{\{1,\ldots,	6\},\{7,8,9\}\}$ to~$\Gamma$
	because
\begin{align*}
        \utilityeqsum_{1}(\Gamma)
	&= v_1(\Gamma) +\cdots
	+v_5(\Gamma)+v_9(\Gamma)
	= (9\cdot 5-3) + 4\cdot ( 9\cdot 2 - 6 )+ (9\cdot 3-5)
	= 112 \\
	&< 113
	= (9\cdot 4-1) + 4\cdot ( 9\cdot 2 - 3 )+ (9\cdot 2-0)
	= v_1(\Delta) + 
	\cdots
	+v_5(\Delta)+v_9(\Delta) \\
	&= \utilityeqsum_{1}(\Delta).
\end{align*}
	Hence, $\Gamma$ is not perfect. 
	\end{example}

\section{Conclusions and Open Problems}

We have proposed to extend the models of altruistic hedonic games due
to
Nguyen \emph{et al.}~\cite{ngu-rey-rey-rot-sch:c:altruistic-hedonic-games}
and  Wiechers and Rothe~\cite{wie-rot:c:stability-in-minimization-based-altruistic-hedonic-games}
to coalition formation games in general. 
We have compared our more general models to altruism in hedonic games and
have motivated our work %
by removing some crucial disadvantages that come with the restriction to hedonic games.
In particular, we have shown that all
degrees of our general altruistic preferences are unanimous
while this is not the case for all
altruistic hedonic preferences.
Furthermore, all our sum-based degrees of altruism fulfill 
two types of monotonicity that
are violated by the corresponding 
hedonic equal- and altruistic-treatment preferences.

We have furthermore studied the common stability notions 
and have initiated a
computational analysis of the associated verification and existence
problems (see Table~\ref{tab:results_SF} for an overview of our results).
We 
also gave
characterizations
for some of the stability notions, using graph-theoretical properties of the underlying network of friends.
For future work, we propose to complete this analysis, close all gaps
between complexity-theoretic upper and lower bounds, and get a full
characterization for all stability notions.

\section*{Acknowledgments}

We thank the anonymous IJCAI'20 reviewers for helpful
comments.  This work was supported in part by DFG grants
RO~1202/14-2 and RO~1202/21-1.
The first and third author have been supported in part by the research
project ``Online Participation'' within the North Rhine-Westphalian
funding scheme ``Forschungskollegs.''

\bibliography{acfgsbib}

\end{document}